\documentclass[reqno,12pt]{article}
\pdfoutput=1
\usepackage{ae}
\usepackage[T1]{fontenc}
\usepackage[ansinew]{inputenc}
\usepackage{mathrsfs}
\usepackage{amsmath}
\usepackage{amssymb}
\usepackage{graphicx}
\usepackage{color}
\definecolor{darkblue}{cmyk}{0.9,0.9,0,0}
\definecolor{darkgreen}{rgb}{0,0.55,0}
\usepackage[colorlinks=true,linkcolor=darkblue,citecolor=darkblue,urlcolor=darkblue]{hyperref}
\usepackage{epsfig}
\usepackage{cite}
\usepackage{mathtools,slashed}
\usepackage{caption}
\usepackage{subcaption}

\usepackage{amsmath,amsfonts,amssymb,amsbsy}
\usepackage{graphicx,color}
\usepackage{graphics}
\usepackage{epsfig}
\usepackage{verbatim}
\usepackage{stmaryrd}
\usepackage{epsf}

\usepackage{amsmath, amssymb,graphicx}
\usepackage{epsfig}
\usepackage{verbatim}
\usepackage{amsfonts}
\usepackage{ulem}
\usepackage{wrapfig}

\numberwithin{equation}{section}

\usepackage{tikz}
\usetikzlibrary{arrows,shapes}
\usetikzlibrary{patterns,fadings}
\usetikzlibrary{decorations.pathmorphing}

\newcommand{\btp}{\begin{tikzpicture}[baseline=0pt,scale=0.9,line width=0.7pt]}
\newcommand{\btpp}{\begin{tikzpicture}[baseline=-5pt,scale=0.25,line width=0.7pt]}
\newcommand{\etp}{\end{tikzpicture}}

\def\bc{\begin{center}}
\def\ec{\end{center}}

\usepackage{epsf}

\newcommand{\be}{\begin{equation}}
\newcommand{\ee}{\end{equation}}
\newcommand{\ba}{\begin{eqnarray}}
\newcommand{\ea}{\end{eqnarray}}
\newcommand{\nn}{{\nonumber}}

\newcommand{\beaa}{\begin{eqnarray}}
\newcommand{\eeaa}{\end{eqnarray}}

\definecolor{darkgreen}{rgb}{0.1,0.7,0.1}

\DeclareFontFamily{OT1}{pzc}{}
\DeclareFontShape{OT1}{pzc}{m}{it}{<-> s * [1.10] pzcmi7t}{}
\DeclareMathAlphabet{\mathpzc}{OT1}{pzc}{m}{it}

\def\({\left(}
\def\){\right)}
\def\[{\left[}
\def\]{\right]}

\def\<{\langle}
\def\>{\rangle}

\def\s*{\ *_{\!\!\!\!\!\!\!\!\!\,_{\,_\text{\scriptsize{sym}}}}}
\def\hs*{\ \hat{*}_{\!\!\!\!\!\!\!\!\!\,_{\,_\text{\scriptsize{sym}}}}}

\begin{document}

\thispagestyle{empty}

\renewcommand{\thefootnote}{\fnsymbol{footnote}}
\setcounter{footnote}{0}
\setcounter{figure}{0}
\begin{center}
$$$$

{\Large\textbf{\mathversion{bold}
Integrable Wilson loops in ABJM: a $Y$-system computation of the cusp anomalous dimension}\par}

\vspace{1.0cm}

\textrm{Diego H. Correa\footnote{correa@fisica.unlp.edu.ar}, Victor I. Giraldo-Rivera\footnote{vigirald@gmail.com} and Mart\'in Lagares\footnote{martinlagares95@gmail.com}}
\\ \vspace{1.2cm}
\footnotesize{

\textit{ Instituto de F\'{\i}sica La Plata, Universidad Nacional de La Plata, \\ C.C. 67, 1900 La Plata, Argentina} \\
\texttt{} \\
}

\par\vspace{1.5cm}

\textbf{Abstract}\vspace{2mm}
\end{center}

\noindent

We study the integrability properties of Wilson loops in the ${\cal N}=6$ three-dimensional Chern-Simons-matter (ABJM) theory. We begin with the construction of an open spin chain that describes the anomalous dimensions of operators inserted along the contour of a 1/2 BPS Wilson loop. Moreover, we compute the all-loop reflection matrices that govern the interaction of spin-chain excitations with  the boundary, including their dressing factors, and we check them against weak- and strong-coupling results. Furthermore, we propose a $Y$-system of equations for the cusped Wilson line of ABJM, and we use it to reproduce the one-loop cusp anomalous dimension of ABJM from a leading-order finite-size correction. Finally, we write a set of BTBA equations consistent with the $Y$-system proposal.

\vspace*{\fill}

\setcounter{page}{1}
\renewcommand{\thefootnote}{\arabic{footnote}}
\setcounter{footnote}{0}

\newpage
\tableofcontents

\newpage

\section{Introduction}
The spectral problem in the three-dimensional ${\cal N}=6$ Chern-Simons-matter theory presented in \cite{Aharony:2008ug,Aharony:2008gk} (usually known as ABJM) is believed to be integrable.
Evidence of integrability was first discovered in the perturbative regime \cite{Minahan:2008hf}, and then in the dual string theory description \cite{Stefanski:2008ik,Arutyunov:2008if,Gromov:2008bz}
(see \cite{Klose:2010ki} for a review).
The underlying integrable structures are similar to those of the four-dimensional ${\cal N}=4$ super Yang-Mills (sYM) theory. There exists, however, a crucial difference: all the exact results that can be extracted using integrability tools are somehow veiled, as they are expressed in terms of a function $h(\lambda)$ of the coupling constant which is in principle unknown.

The resemblance of the integrability calculation for the slope function of ABJM with the matrix model integral that gives the expectation value of certain Wilson loops led the authors of \cite{Gromov:2014eha} to make a conjecture for the unknown function $h(\lambda)$. Moreover, this was later generalized to the ABJ theory in \cite{Cavaglia:2016ide}. A direct derivation of $h(\lambda)$, as it was several times suggested, could be achieved from an integrability-based computation of the ABJM's bremmstrahlung function, as done in the ${\cal N}=4$ sYM theory \cite{Gromov:2012eu}. Exactly the {\it same} magnitude is explicitly obtained from localization \cite{Lewkowycz:2013laa,Bianchi:2014laa,Bianchi:2017svd,Bianchi:2018scb}. Therefore, the comparison of these two results  would provide a derivation of the unknown function $h(\lambda)$.

A starting point for this program could then be  to determine whether ABJM Wilson loops set integrable open boundary conditions for insertions along the contour. This longstanding problem was schematically discussed in \cite{Drukker:2019bev} and will be revisited here.  The fact that the perturbative anomalous dimensions of operators inserted within Wilson loops can be described by integrable open spin chain was first observed in \cite{Drukker:2006xg} for the case of the 1/2 BPS Wilson loop in ${\cal N}=4$ sYM\footnote{As shown in \cite{Correa:2018fgz}, the perturbative anomalous dimensions of operators inserted within the ordinary Wilson loops are also described by an integrable open spin chain.}. We will study a generalization of this result to the case of 1/2 BPS Wilson loops in ABJM. To that aim we shall begin with the construction of an appropriate vacuum reference state for the open spin chain. As we will see, a large R-charge insertion in the Wilson loop, which will play the role of the vacuum state of the spin chain, can be at most 1/6 BPS. This is inferred from the existence of 1/6 BPS rotating folded strings 
in the dual $AdS_4\times {\mathbb{CP}^3}$ background.
 Magnon excitations propagate along the spin chain vacuum state and their bulk scattering and reflection matrices are determined upon symmetry considerations.  Following this argument, the boundary reflection matrix was proposed in \cite{Drukker:2019bev} up to an overall dressing phase. The fact that the (boundary) Yang-Baxter equations are satisfied is regarded as an indication of the integrability of the system. 
We will obtain the corresponding dressing phase non-perturbatively 
by solving an appropriate crossing equation and by demanding consistency with explicit weak- and strong-coupling computations.
Interestingly, this phase will be radically different for the two types of magnons that can propagate in the alternating spin chain.

In the case of the ${\cal N}=4$ sYM theory, the integrability of insertions within Wilson loops was successfully applied to the computation of the cusp anomalous dimension via a Boundary Thermodynamic Bethe Ansatz (BTBA) in \cite{Correa:2012hh,Drukker:2012de}. 
In this paper we will extend these ideas to compute the cusp anomalous dimension $\Gamma_{\rm cusp}$ of ABJM. More precisely, we will take a Wilson loop with a cusp and we will consider the insertion of a spin-chain vacuum at the position of the cusp. We will propose a $Y$-system of equations and we will compute from it the finite-size correction to the corresponding vacuum energy as a function of its length. Eventually, the cusp anomalous dimension is obtained by evaluating the vacuum energy when its length is taken to be zero. As a verification of our proposal, we will show that it consistently reproduces the one-loop value of $\Gamma_{\rm cusp}$.
In the small cusp angle limit, $\Gamma_{\rm cusp}$ gives the bremmstrahlung function. Thus, if the  $Y$-system equations that we propose could be exactly solved in this limit, it would provide a direct derivation of the interpolating function $h(\lambda)$ of ABJM.

The paper is organized as follows. We begin in Section \ref{sec: review spin chains} with a short overview of the main features of integrable spin chains in ABJM, and in Section \ref{review WL} we review the properties of the 1/2 BPS Wilson loop of ABJM. In Section \ref{sec: open spin chain} we discuss the construction of integrable open spin chains to describe the anomalous dimensions of operators inserted within the 1/2 BPS Wilson loop of ABJM. Section \ref{sec: crossing} is devoted to the computation of the boundary dressing phases that describe the reflection of magnons in the open spin chain. In Section \ref{sec: Gamma cusp} we give our proposal for the $Y$-system and BTBA equations that describe the finite-size corrections to the vacuum energy of the cusped Wilson line of ABJM, and we use such equations to reproduce the one-loop cusp anomalous dimension of ABJM from a leading-order finite-size correction. Finally, we give our conclusions in Section \ref{sec: conclusions}. We include two Appendices that complement the results presented in the main body of the paper. In Appendix \ref{app: ABJM conventions} we give our conventions for the ABJM theory, while in Appendix \ref{app: String theory description} we propose a dual string for the vacuum state of the open spin chain.


\section{Integrable spin chains in ABJM}
\label{sec: review spin chains}

Since the seminal work of Minahan and Zarembo \cite{Minahan:2002ve}, integrability has played a mayor role in the computation of anomalous dimensions of operators within the framework of the AdS/CFT correspondence. More precisely, the matrix of anomalous dimensions of single-trace operators was identified, first in ${\cal N}=4$ sYM \cite{Minahan:2002ve} and then in ABJM \cite{Minahan:2008hf}, with the hamiltonian of an integrable spin chain. In that context, the vacuum state of the spin chain picture corresponds to a protected operator of the corresponding gauge theory. In particular, in the  ABJM theory (see Appendix \ref{app: ABJM conventions} for conventions) the corresponding BPS operator is
\begin{equation}
    \label{periodic vacuum d=3}
    {\rm Tr} \left[ \left( C_1 \bar{C}^2 \right)^{\ell} \right]\,,
\end{equation}
with $\ell \in \mathbb{N}$.  
The presence of the trace in \eqref{periodic vacuum d=3} implies that the spin chain is \textit{periodic}. When considering excited states a novel feature appears in the ABJM theory with respect to the ${\cal N}=4$ sYM case. More specifically,  in the ABJM picture one can construct an excited state either by replacing a $C_1$ or a $\bar{C}^2$ field with an impurity. Therefore, excitation waves (i.e. \textit{magnons}) can be of two types,
\begin{align}
    \text{type\ } A \text{\ magnons:} & \qquad (C_3,C_4|\bar\psi^2_+,\bar\psi^2_-)\,,
    \\
     \text{type\ } B \text{\ magnons:} & \qquad
     (\bar C^3,\bar C^4|\psi_1^+,\psi_1^-)\,.
\end{align}
From the above, we say that the periodic spin chain of ABJM is \textit{alternating}, as there are two distinct sites along which impurities can propagate.

Taking into account the symmetries of the spin-chain system is crucial, as they are used to determine the scattering properties of the magnons. While in ${\cal N}=4$ sYM the vacuum state has $SU(2|2)^2$ invariance, in ABJM the corresponding symmetry is reduced to just one copy of $SU(2|2)$. In the general case, one can consider bound states of $Q$ magnons propagating along the chain, with the simplest $Q=1$ case being a single-particle state.
Each $Q$-magnon state 
accommodates in  a 
representation of the $SU(2|2)$  symmetry group\cite{Gaiotto:2008cg}, 
and each of those representations is labelled by four coefficients $({\sf a},{\sf b},{\sf c}, {\sf d})$ that characterize the action of the fermionic generators over the corresponding states. As an example, for $Q=1$ (i.e. the fundamental representation) one has
\begin{equation}
\begin{array}{ll}
   Q_{a}^{\alpha}|\phi^b\rangle
 =  {\sf a}\,\delta_{a}^{b}|\psi^\alpha\rangle,  &  Q_{a}^{\alpha}|\psi^{\beta}\rangle = {\sf b}\, \epsilon^{\alpha\beta}\epsilon_{ab}|{\phi}^{b}\rangle , \\
  S_{\alpha}^{a}|\phi^{b}\rangle
 = {\sf c}\,\epsilon_{\alpha\beta}\epsilon^{ab}|\psi^{\beta}\rangle,    & 
  S_{{\alpha}}^a|\psi^{\beta}\rangle
= {\sf d}\, \delta_{\alpha}^{\beta}|\phi^{a}\rangle ,
\end{array}
\label{QSonfundamental}
\end{equation}
with $a=1,2$ and $\alpha=3,4$. In the general case, the labels depend on the magnon momentum and on an unknown interpolating function $h(\lambda)$ as\footnote{We shall focus on the ABJM case, in which both gauge groups have equal ranks.}
\begin{equation}
{\sf a}=\sqrt{\frac{h(\lambda)}{Q}}\eta,\quad
{\sf b}=\sqrt{\frac{h(\lambda)}{Q}}\frac{i\zeta}{\eta}(\tfrac{x^{+}}{x^{-}}-1),\quad
{\sf c}=-\sqrt{\frac{h(\lambda)}{Q}}\frac{\eta}{\zeta x^{+}},\quad
{\sf d}=-\sqrt{\frac{h(\lambda)}{Q}}\frac{x^{+}}{i\eta}(\tfrac{x^{-}}{x^{+}}-1)\,,
\label{14:abcd}
\end{equation}
with 
\begin{equation}
x^{+}+\frac{1}{x^{+}}-x^{-}-\frac{1}{x^{-}}=\frac{iQ}{h(\lambda)},
\qquad \frac{x^+}{x^-}= e^{ip},\qquad
\eta(p,\zeta) = \zeta^{\frac 12} e^{\frac{i p}4}\sqrt{i(x^--x^+)}\,,
\label{14:mass-shell_1}
\end{equation}
and where $p$ is the magnon momentum and $\zeta$ is a phase. The unknown function is also present in the dispersion relation of magnons,
\begin{equation}
\label{magnon energy}
E(p)=\frac{1}2\sqrt{Q^2+16 h^2(\lambda)\sin^2(p/2)}\,,
\end{equation}
and it therefore percolates in all the results obtained with integrability techniques.

Interestingly, as proven in \cite{Beisert:2005tm}
the $SU(2|2)$ symmetry of the spin chain is enough to bootstrap the all-loop $2\to2$ scattering matrix of the problem, up to an overall coupling-dependent \textit{dressing phase}. Moreover, a non-perturbative computation of the latter was achieved in \cite{Beisert:2006ez}. Being an integrable system, the previous results determine completely the full scattering matrix of the theory. To be more specific, the $SU(2|2)$ symmetry of the reference state fixes the scattering matrices of type $A$ and type $B$ magnons to be \cite{Ahn:2008aa}
\begin{equation}
\label{S matrix ABJM}
\begin{array}{c}
    S^{AA}(x_1,x_2) = S^{BB}(x_1,x_2) = S_0(x_1,x_2)\hat S(x_1,x_2)\,,
    \\
    S^{AB}(x_1,x_2) = S^{BA}(x_1,x_2) = \tilde S_0(x_1,x_2)\hat S(x_1,x_2)\,,
\end{array}    
\end{equation}
 where $\hat S(x_1,x_2)$ is the $SU(2|2)$-invariant matrix\footnote{We use $\hat S^{11}_{11}(x_1,x_2)=1$.} given in \cite{Arutyunov:2008zt} while $S_0(x_1,x_2)$ and $\tilde S_0(x_1,x_2)$ are the dressing phases. The scalar factors can be fixed by demanding crossing symmetry to be \cite{Ahn:2008aa,Chen:2019igg}
\begin{equation}
\label{S dressing phases}
\tilde S_0(x_1,x_2) = \sqrt{\frac{x_1^-}
{x_1^+}} \, \sqrt{\frac{x_2^+}
{x_2^-}} \, \sigma(x_1,x_2) \,, \qquad
S_0(x_1,x_2) = \frac{x_1^+-x_2^-}{x_1^--x_2^+} \,
\frac{1-\frac{1}{x_1^+ x_2^-}}{1-\frac{1}{x_1^- x_2^+}} \, \tilde S_0(x_1,x_2)  \,,
\quad 
\end{equation}
where $\sigma(x_1,x_2)$ is the BES dressing factor \cite{Beisert:2006ez}.


\section{Supersymmetric Wilson loops in ABJM}
\label{review WL}

As symmetries play a central role in fixing the reflection matrix of magnons, we consider convenient to provide here a short review of how supersymmetric Wilson loops are constructed in the ABJM theory. 

In ${\cal N}=4$ sYM the construction of 1/2 BPS Wilson loops is achieved simply by considering a straight (or circular) Wilson loop with a constant coupling to the scalar fields of the theory. The generalization of this idea to the ABJM case has proven to be more subtle, as adding a constant coupling to the scalars gives an operator which is at most 1/6 BPS \cite{Drukker:2008zx}. The construction of 1/2 BPS Wilson loops in ABJM was successfully achieved in \cite{Drukker:2009hy} by taking into account the holonomy of a $U(N|N)$ \textit{superconnection}\footnote{This is a particular case of the results presented in \cite{Drukker:2009hy}, where the authors studied the more general picture in which both gauge groups can have different ranks.}. To be more specific, let us consider a Wilson loop defined as
\be
\label{general WL}
W = \frac{1}{2N}{\rm Tr} \left[  {\cal W}(\tau_1,\tau_2) \right]
 :=
\frac{1}{2N} {\rm Tr} \left[
 {\cal P}\exp\left( {i \int_{\tau_1}^{\tau_2}
 {\cal L}(\tau) d\tau}\right)
\right]\,,
\ee
with\footnote{We use conventions in which the usual factor $\sqrt{\frac{2\pi}k}$ has been absorbed into the scalars and fermions of the theory.}
\be
{\cal L} =\begin{pmatrix}
A_\mu \dot{x}^\mu-i|\dot{x}|{M}^{I}_J C_I \bar{C}^J & -i |\dot{x}| \eta^{\alpha}_{I} \bar{\psi}^{I}_{\alpha}
\\
 -i |\dot{x}| \psi^{\alpha}_{I} \bar{\eta}^{I}_{\alpha} & \hat A_\mu \dot{x}^\mu-i|\dot{x}|{M}^{I}_J \bar{C}^J C_I
\end{pmatrix}\,,
\ee
and where $M^I_J$ and $\eta^{\alpha}_{I}$ are Grassmann-even couplings. It turns out that the condition 
\be
\label{restrivctive susy constraint}
\delta_{\rm SUSY} {\cal L}=0 \,,
\ee
is too restrictive to  have a 1/2 BPS operator (see Appendix \ref{app: ABJM conventions} for the supersymmetry transformations of ABJM). Instead, one should relax the constraint \eqref{restrivctive susy constraint} by demanding that supersymmetry translates into a super-gauge transformation for ${\cal L}$, i.e. 
\be
\delta_{\rm SUSY} {\cal L}= {\cal D}_{\tau} \Lambda := \partial_{\tau} \Lambda +i \{ {\cal L}, \Lambda ] \,,
\ee
where $\Lambda$ is some supermatrix. With this choice, under a finite supersymmetry transformation one gets\footnote{To construct a Wilson loop invariant under these finite transformations one has to be careful with the choice of boundary conditions. While a straight contour is compatible with taking the trace in the definition \eqref{general WL}, for a closed contour with periodic boundary conditions one should instead consider the supertrace \cite{Cardinali:2012ru}.}
\be
\label{WL finite transformation}
{\cal W} (\tau_1,\tau_2) = {\cal P}\exp\left( {i \int_{\tau_1}^{\tau_2}
 {\cal L}(\tau) d\tau}\right) \to U^{-1} (\tau_1) {\cal W} (\tau_1,\tau_2) U (\tau_2) \,,
\ee
with $U(\tau)= \exp[i \Lambda (\tau)]$. In this framework, one can see that by taking the orientations specified by
\be
x^\mu =(\tau,0,0)\,,\quad M^I_{J} = -\delta^I_J +2\delta^I_1\delta^1_J\,,\quad
\eta_I^\alpha  = \eta \, \delta_I^1\delta^\alpha_+\,,\quad
\bar\eta^I_\alpha  = \bar\eta \, \delta^I_1\delta_\alpha^+\,,
\label{WLchoice}
\ee
the resulting Wilson line is 1/2 BPS provided
\be
\eta\bar\eta = -2i \,.
\ee
In particular, with the choice \eqref{WLchoice}, the Wilson loop \eqref{general WL} is invariant under the supersymmetry transformations \eqref{susy ABJ(M) 1} for arbitrary non-vanishing values of $\bar{\Theta}^{1J}_+ $ and $\bar{\Theta}^{IJ}_-$  with  $I,J\neq 1$. The corresponding $\Lambda$ matrix is
\be
\label{lambda straight WL}
\Lambda = \left(
\begin{array}{cc}
0 & g_1
\\
\bar g_2 & 0
\end{array}\right)\,,
\ee
with\cite{Cardinali:2012ru}
\be 
g_1 = 2 \, \eta \, \bar{\Theta}^{1I} C_I \,,
\qquad
\bar g_2 = \epsilon_{1IJK} \, \bar{\eta} \, \bar{\Theta}^{IJ} \bar{C}^K \,.
\ee

It will prove useful to note that the Wilson loop defined by the parametrization \eqref{WLchoice} is invariant under an $SU(1,1|3)$ subgroup of the original $OSp(6|4)$ symmetry of ABJM. For a detailed discussion on the corresponding representation theory see \cite{Bianchi:2017ozk,Bianchi:2020hsz}.

Finally, let us take a Wilson loop with a cusp described by an angle $\theta$, given for example by the parametrization
\be
x^0=0 \,, \qquad x^1= s \cos \frac{\theta}{2} \,, \qquad x^2= |s| \sin \frac{\theta}{2}\,, \qquad -\infty< s < \infty \,.
\ee
We will focus on a geometric cusp and we will not consider a cusp in the internal space orientation, described by the couplings $M^I_J$ and $\eta_I^{\alpha}$. As is well known from the renormalization theory of Wilson loops \cite{Polyakov:1980ca,Dotsenko:1979wb,Brandt:1981kf,Korchemskaya:1992je}, the presence of a cusp in the contour introduces divergences that can not be absorbed with a redefinition of the couplings of the theory. More precisely, when regularizing such divergences one gets
\be
\langle W^{\rm ren}(\theta) \rangle= Z_{\rm cusp} (\theta) \langle W(\theta) \rangle \,,
\ee
where $\langle W^{\rm ren}(\theta) \rangle$ is the renormalized v.e.v. of the Wilson loop and $Z_{\rm cusp} (\theta)$ is the corresponding renormalization factor. Therefore, one can naturally define a \textit{cusp anomalous dimension} as
\be
\Gamma_{\rm cusp} (\theta) = \mu \frac{\partial Z_{\rm cusp} (\theta)}{\partial \mu} \,,
\ee
where $\mu$ is the renormalization scale of the theory. A perturbative computation of $\Gamma_{\rm cusp} (\theta)$ was performed in \cite{Griguolo:2012iq}, giving\footnote{Taking the limit of $\theta$ large and imaginary in the perturbative results of \cite{Griguolo:2012iq} gives
\be
\label{perturbative light like gamma cusp}
\Gamma_{\rm cusp, light} = \lambda^2 + \mathcal{O}(\lambda^3) \,,
\ee
for Wilson loops with light-like cusps, in accordance with the all-loop proposal of \cite{Gromov:2014eha,Gromov:2008qe}. Recently, a geometric approach for the computation of $\Gamma_{\rm cusp, light}$ was studied in \cite{Henn:2023pkc}.}
\be
\label{perturbative gamma cusp}
\Gamma_{\rm cusp} (\theta) = - \lambda \left( \frac{1}{\cos\frac{\theta}{2}} -1 \right) + \mathcal{O}(\lambda^2) \,, 
\ee
at leading order. In the following sections we will reproduce the above result using an integrability approach.


\section{Wilson loop's open spin chain}
\label{sec: open spin chain}

In order to compute the cusp anomalous dimension $\Gamma_{\rm cusp} (\theta)$ from an integrability approach we should first study the description of insertions within Wilson loops in terms of open spin chains. We will devote this section to such goal.

To begin with, we should identify which insertion could serve as the vacuum state of the Wilson loop spin-chain system. Following the insight obtained from the ${\cal N}=4$ sYM case, in ABJM one expects that a vacuum state with large $R$ charge should be dual to a BPS string ending on the Wilson loop's contour at the boundary of $AdS_4$ and with large angular momentum in the coordinates of the $\mathbb{CP}^3$ compact space. As shown in the Appendix \ref{app: String theory description}, one can construct a 1/6 BPS string with those properties which is invariant under a $SU(1|2)$ supersymmetry. We will therefore search for a vacuum state with the same supersymmetry.

Naively, one might expect that 
\be
{\cal D} := \left(
\begin{array}{cc}
(C_1 \bar C^2)^\ell & 0
\\
0 &  (\bar C^2 C_1)^\ell
\end{array}\right),
\label{vacuum12}
\ee
could be the vacuum state we are looking for,
as it shares some supersymmetry with the 1/2 BPS Wilson loop. However, 
the total operator, i.e. the Wilson loop with the operator \eqref{vacuum12} inserted at a position $\tau$, is not supersymmetric. Because the path-ordered exponential ${\cal W}(\tau_1,\tau_2)$ is covariant rather than invariant under supersymmetry, studying the supersymmetry transformations of insertions within Wilson loops is a bit more subtle \cite{Bianchi:2020hsz}.
To be more specific, let us define  ${\cal O}_W$ as the insertion of a generic operator ${\cal O}$ at the point $\tau$, i.e.
\begin{equation}
\label{WL with general insertion}
{\cal O}_{W}(\tau) := \frac{1}{2N} {\rm Tr} \left[{\cal P} {\cal W} (-\infty,\tau) \, \mathcal{O} (\tau) \, {\cal W} (\tau,\infty) \right]\,.
\end{equation}
In order to consider the transformation \eqref{WL finite transformation} of the complete operator ${\cal O}_{W}$ under supersymmetry it is instructive to introduce a \textit{covariant} supersymmetric transformation \cite{Gorini:2022jws} as
\begin{equation}
\label{covariant susy transformation}
\delta^{\rm cov} {\cal O} :=\delta {\cal O} - i [{\cal O}, \Lambda] \,.
\end{equation}
In this context, we will say that an insertion is supersymmetric if
\begin{equation}
\label{BPS condition for O}
\delta^{\rm cov} {\cal O} =0\,.
\end{equation}
Therefore, we see that despite satisfying $\delta{\cal D} =0$ the operator ${\cal D}_W$ is not supersymmetric, because the insertion does not commute with the $\Lambda$ matrix given in \eqref{lambda straight WL}. Instead, it is straightforward to verify that for
\be 
{\cal T_+} = 
\left(
\begin{array}{cc}
C_1 \bar C^2& -\frac{\eta}{2}\bar\psi^2_+
\\
0 &  \bar C^2 C_1
\end{array}\right)\,,
\label{susyinsertion}
\ee
the condition \eqref{BPS condition for O} is met. An arbitrary power of this operator will be equally BPS and provides an insertion with a large amount of the corresponding R-charge
\be 
{\cal T}_+^{\ell} = 
\left(
\begin{array}{cc}
(C_1 \bar C^2)^\ell& -\frac{\eta}{2}
\sum_k (C_1 \bar C^2)^k\bar\psi^2_+
(\bar C^2 C_1)^{\ell-k-1}
\\
0 &  (\bar C^2 C_1)^\ell
\end{array}\right)\,.
\label{susyinsertionlarge}
\ee

Although protected, we shall not consider \eqref{susyinsertionlarge} as the reference state to formulate a Bethe Ansatz. Its off-diagonal block looks more like a one-impurity state (with zero momentum) than a vacuum state. Moreover, as we shall see next, the operator \eqref{susyinsertionlarge} can be regarded as a descendant when considering the covariant action of the supersymmetry transformations on a certain insertion within the Wilson line.

A more appropriate alternative to play the role of a Bethe Ansatz reference state turns out to be the off-diagonal insertion
\be 
{\cal V}_\ell = 
\left(
\begin{array}{cc}
 0 & (C_1 \bar C^2)^\ell C_1 
\\
0 &  \bar 0
\end{array}\right)\,.
\label{susyinsertion0}
\ee
This operator is invariant under the supersymmetries generated by
$\bar{\Theta}^{13}_+$ and $\bar{\Theta}^{14}_+$, for which the $\Lambda$ matrix is given by
\be 
g_1 = 2{\eta}\left(\bar{\Theta}^{13}_+ C_3 + \bar{\Theta}^{14}_+ C_4\right)\,,
\qquad
\bar g_2 = 0\,.
\ee
Therefore, the insertion \eqref{susyinsertion0} breaks the $SU(1,1|3)$ symmetry of the Wilson loop to $SU(1|2)$, as expected in view of the results coming from the string theory side of the AdS$_4$/CFT$_3$ duality. Moreover, when acting with the supersymmetry transformation generated by $\bar{\Theta}^{34}_-$ on ${\cal V}_\ell$ one precisely obtains ${\cal T}_+^{\ell+1}$. Consequently, in what follows we shall consider \eqref{susyinsertion0} as our Bethe Ansatz reference state.

Having identified a suitable vacuum state, we can now turn to the analysis of the impurities that can propagate along the spin chain. From the inspection of the operator \eqref{susyinsertion0}, one can see that the excited states that  propagates over such vacuum are a straightforward generalization of the type $A$ and type $B$ magnons of the periodic spin chain. Moreover, the S-matrix is the same $SU(2|2)$-invariant matrix that governs the scattering of magnons in the periodic case, as the presence of the boundary (i.e. the Wilson loop) does not affect the bulk interactions.

In addition to the bulk scattering, for open spin chains one also has to take into account the reflection of magnons against the boundary, which is characterized by a \textit{reflection matrix}. Let us focus now on its computation. As discussed above, there is a $SU(1|2)$ residual symmetry preserved by the Wilson loop
\eqref{WLchoice} with the insertion \eqref{susyinsertion0}. Following \cite{Hofman:2007xp,Correa:2008av}, one can use this symmetry to constrain the boundary reflection matrix of magnon excitations. The action of the right reflection matrix over the quantum numbers of a fundamental representation can be taken such that 
$(p,\zeta) \mapsto (p,-\zeta)$.
The most general reflection matrix ${\mathbf R}$ would in principle allow for the mixing of magnons of type $A$ and $B$,
\begin{equation}
{\mathbf R} = \left(
\begin{array}{cc}
R_{A}     &  \tilde{R}_{A} \\
\tilde{R}_{B}     &  R_{B}
\end{array}
\right)\,,
\end{equation}
where $R_{A/B}$ indicates the reflection of a type $A/B$ magnon into a type $A/B$ magnon. On the contrary, $\tilde{R}_{A/B}$ indicates  the reflection of a type $A/B$ magnon into a type $B/A$ magnon. The $SU(1|2)$ residual symmetry constrains the form of each of the blocks to be \cite{Drukker:2019bev}
\begin{align}
R_{A} & = R_{A}^0 \, {\rm diag}(1,1,e^{-ip/2},-e^{ip/2}) \,,
\label{RAA}
\qquad
\tilde{R}_{A}  =  \tilde{R}_{A}^0 \, {\rm diag}(1,1,e^{-ip/2},-e^{ip/2}) \,,
\\
R_{B} &  = R_{B}^0 \, {\rm diag}(1,1,e^{-ip/2},-e^{ip/2}) \,,
\qquad
\tilde{R}_{B} =  \tilde{R}_{B}^0 \, {\rm diag}(1,1,e^{-ip/2},-e^{ip/2}) \,,
\end{align}
where $R_{A}^0,R_{B}^0,  \tilde{R}_{A}^0,  \tilde{R}_{B}^0$ are dressing factors that can not be fixed with symmetry arguments.  With a reflection matrix of this form, the boundary Yang-Baxter equation is not satisfied unless $R_{A}^0 = R_{B}^0 = 0$ or $ \tilde{R}_{A}^0 = \tilde{R}_{B}^0 = 0$. The weak-coupling analysis we will present in the next section shows that, at least perturbatively, the reflection at the boundaries does not mix type $A$ and type $B$ magnons. In the following we will consider the validity of $ \tilde{R}_{A}^0 = \tilde{R}_{B}^0 = 0$
at all-loop as a working assumption.

\section{Crossing symmetry and boundary dressing factors}
\label{sec: crossing}

Even in the case of  no mixing between different types of magnons at the boundary, the reflection matrix is only known up to two boundary dressing factors ${R}_{A}^0(p)$ and ${R}_{B}^0(p)$. In this section we will focus on their computation, using the standard constraints coming from boundary crossing-unitary conditions. Among the many solutions to the crossing equations we shall single the ones that are consistent with explicit weak- and strong-coupling computations.

\subsection{Crossing equation}
\label{sec: crossing eq}

We will follow the ideas of \cite{Hofman:2007xp} to derive the boundary crossing equation. More specifically, we will consider the reflection of a singlet state against the boundary, and we will obtain the boundary crossing equation by demanding that such reflection must be trivial.

Let us start with the construction of the singlet state, whose defining property is its trivial interaction with any other particles. Taking this into account, one should look for a state whose quantum numbers coincide with those of the vacuum. Let us recall that in ABJM the spin chain has a $U(1)_{\rm extra}$ symmetry under which the fields $\bar{C}^1,\bar{C}^2,C_3,C_4$ have charge +1 and the fields $C_1,C_2,\bar{C}^3$ and $\bar{C}^4$ have charge -1 \cite{Klose:2010ki}. Therefore, 
we should search for a singlet state with the same $U(1)_{\rm extra}$ 
charge  as the vacuum. With this in mind, we will consider
\begin{equation}
\label{singlet}
|1_{AB} \rangle (p, \bar{p}) = \epsilon_{ab} |\phi^a_A (p) \phi^b_B (\bar{p}) \rangle + \kappa \, \epsilon_{\alpha \beta} |\psi^{\alpha}_A (p) \psi^{\beta}_B (\bar{p}) \rangle \,.
\end{equation}
In \eqref{singlet} we have $\kappa \in \mathbb{R}$ and the crossing transformation $\bar{p}$ is defined such that
\begin{equation}
\label{crossing transformation}
p \to -p  \quad \text{and} \quad E \to -E \qquad \Leftrightarrow \qquad x^{\pm} \to \frac{1}{x^{\pm}} \quad \text{and} \quad \bar{\zeta} \to \zeta \frac{x^+}{x^-} \,.
\end{equation}
For \eqref{singlet} to be a singlet state we have to further demand its invariance under all the $SU(1|2)$ generators, which implies
\begin{equation}
\label{kappa}
\kappa = - \frac{i x^-}{(x^- - x^+) \, \zeta } \, \eta(x^+,x^-,\zeta) \, \eta \left( \tfrac{1}{x^+},\tfrac{1}{x^-},\zeta \tfrac{x^+}{x^-} \right) \,,
\end{equation}
where $\eta$ was defined in \eqref{14:mass-shell_1}.

We will obtain the crossing equation by demanding that, under a sequence of right and left reflections, the singlet state \eqref{singlet} remains invariant, i.e. 
\begin{equation}
\label{right reflection singlet-invariance condition}
R_{A}^L (-p) S^{AB} (-p, \bar{p}) R_{B}^L (-\bar{p}) R_{A}^R (p) S^{AB} (p, -\bar{p}) R_{B}^R (\bar{p}) |1_{AB} \rangle (p, \bar{p}) = |1_{AB} \rangle (p, \bar{p}) \,,
\end{equation}
where we have introduced the labels $R$ and $L$ to refer to the right- and left- reflection matrices, respectively. Then, recalling that the action of a reflection is such that
\begin{equation}
\label{ref transformation}
p \to -p \qquad  \text{and} \qquad E \to E \qquad \Leftrightarrow \qquad x^{\pm} \to -x^{\mp} \,,
\end{equation}
we get that for the right reflection
\begin{equation}
\label{crossing condition}
R_{A}^R (p) S^{AB} (p, -\bar{p}) R_{B}^R (\bar{p}) |1_{AB} \rangle (p, \bar{p}) = r(p) |1_{AB} \rangle (-p, -\bar{p}) \,,
\end{equation}
where the reflection phase is
\begin{equation}
\label{reflection phase singlet}
r(p) = \frac{\frac{1}{x^-}+x^-}{\frac{1}{x^+}+x^+} \sigma \left( p, -\bar{p} \right)   R_{A}^0 (p) R_{B}^0 \left( \bar{p} \right) \,.
\end{equation}
Parity invariance demands that the reflection at the left boundary results in the same reflection phase \cite{Hofman:2007xp}. Therefore,
\begin{equation*}
\label{right reflection singlet-invariance condition-2}
R_{A}^L (-p) S^{AB} (-p, \bar{p}) R_{B}^L (-\bar{p}) R_{A}^R (p) S^{AB} (p, -\bar{p}) R_{B}^R (\bar{p}) |1_{AB} \rangle (p, \bar{p}) = r(p)^2 |1_{AB} \rangle (p, \bar{p}) \,.
\end{equation*}
By demanding \eqref{right reflection singlet-invariance condition} we get
\begin{equation}
\label{crossing eq-1}
r(p)^2=1 \,.
\end{equation}
Between the solutions $r(p)=1$ and $r(p)=-1$, we will see in the next section that only the latter is 
compatible with weak-coupling results. Consequently,
\begin{equation}
\label{crossing eq-2}
R_{A}^0 (p) R_{B}^0 \left( \bar{p} \right) = - \frac{\frac{1}{x^+}+x^+}{\frac{1}{x^-}+x^-}  \frac{1}{\sigma \left( p, -\bar{p} \right)} \,.
\end{equation}
Finally, we have to impose also the unitarity constraints \cite{Ghoshal:1993tm}
\begin{equation}
\label{unitarity constraints}
\begin{aligned}
R_{A}^0(-p) R_{A}^0(p) &=1 \,, \\
R_{B}^0(-p) R_{B}^0(p) &=1 \,.
\end{aligned}
\end{equation}

\subsection{Weak-coupling analysis}
\label{sec: weak coupling}

In order to get insight towards the construction of an all-loop solution of the crossing equation, let us focus now on the weak-coupling expansion of the dressing phases $R_{A}^0$ and $R_{B}^0$. 

Let us begin by studying an $SU(2)$ scalar sub-sector for the odd sites of the chain, where type $A$ impurities can be allocated. We will consider states of the form
\begin{equation}
\label{A impurity states}
| C_{I_0} , C_{I_1},\cdots C_{I_{\ell}}\rangle := \sqrt{2} \left( \frac{2k}{N} \right)^{\ell+\frac{1}{2}} \left( \begin{array}{cc} 0 & C_{I_0}\bar{C}^2C_{I_1}\bar{C}^2\cdots \bar{C}^2C_{I_{\ell}}
\\ 
0 & 0 \end{array} \right) , 
\end{equation}
where $k$ is the Chern-Simons level, $N$ is the number of colors and the $ C_{I_n}$ fields can be either $C_1$ or $C_3$. The overall constant in \eqref{A impurity states} has been included to get a trivial normalization in the tree-level contribution to the two-point functions.

We shall now turn to the computation of the Hamiltonian ${\bf H}^{A}$, which governs the quantum 
dynamics of the states $| C_{I_0} , C_{I_1},\cdots C_{I_{\ell}}\rangle$. Let us recall that ${\bf H}^{A}$ is given by the perturbative mixing matrix of anomalous dimensions of the corresponding operators, which can be computed from the correlator between an operator \eqref{A impurity states} inserted at $\tau_2$ and a conjugate operator inserted at $\tau_1$.
It is useful to distinguish between the two types of Feynman diagrams contributing to the mixing matrix of anomalous dimensions: those in which the contribution of the Wilson line is trivial and those including  propagators from the Wilson line. The former give rise to the \textit{bulk} Hamiltonian ${\bf H}_{\rm bulk}^{A}$. The latter, in contrast,
specify the \textit{boundary} Hamiltonian ${\bf H}_{\rm bdry}^{A}$.

Since for the moment we are just interested in the computation of the reflection matrix, we can ignore the right boundary and focus on the left one. 
Therefore, we will take the $\ell \to \infty$ limit and we will deal with a semi-infinite chain whose only boundary is at the left.
As discussed in Section \ref{sec: open spin chain}, ${\bf H}_{\rm bulk}^{A}$ should not be different from the periodic spin-chain Hamiltonian, and so at two-loops we have
\begin{equation}
\label{bulk hamiltonian}
{\bf H}_{\rm bulk}^{A} = \lambda^2 \sum_{n=0}^{\infty} ({\bf 1}-{\bf P}_{n,n+1}) \,,
\end{equation}
where ${\bf P}_{n,n+1}$ is the permutation operator between fields at the sites $n$ and $n+1$. 

\begin{figure}
\centering
\includegraphics[scale=0.7]{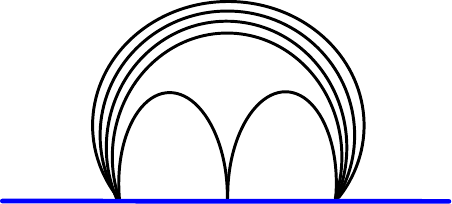}
\hspace{0.5cm}
\includegraphics[scale=0.7]{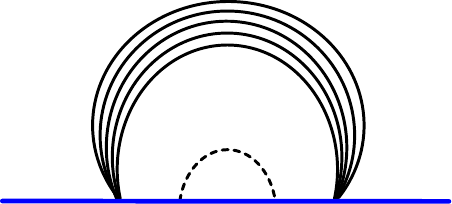}
\caption{Feynman diagram contributing to ${\bf H}_{\rm bdry}^{A}$ at 1-loop. Solid and dashed lines represent scalar and fermionic  propagators respectively. The blue line is the Wilson loop.}
\label{calzonfermion}
\end{figure}

Therefore, we just have to compute the diagrams that give ${\bf H}_{\rm bdry}^{A}$. As customary when evaluating Feynman diagrams in $d = 3-2\epsilon$, the anomalous dimensions can be read from the residue in the $1/\epsilon$ divergences. The only diagrams that contribute at 1-loop order are the ones depicted in Fig. \ref{calzonfermion}. As anticipated, there is no mixing between type $A$ and $B$ impurities and the action of ${\bf H}_{\rm bdry}^{A}$ is diagonal. When computing the divergence of these two one-loop diagrams one gets
\begin{equation}
\label{divergence alpha}
\frac{\lambda}{(\tau_2-\tau_1)^{2\ell+1}} \frac{1}{\epsilon}
\frac{1}{2}\left(M^{J_0}_{I_0} -\delta^{J_0}_{I_0} \right)
\delta^{J_1}_{I_1} \cdots \delta^{J_{\ell}}_{I_{\ell}}
+ \mathcal{O}(\epsilon) \,.
\end{equation}
From \eqref{divergence alpha} we see that the diagram for the boundary scalar interaction depends on the flavor of left-most field, and the total contribution is non-vanishing when the first site is occupied by a $C_3$. 
With these results we get that the action  of ${\bf H}_{\rm bdry}^{A}$ is of the form
\begin{equation}
\label{bdry hamiltonian}
{\bf H}_{\rm bdry}^{A} =  ( \lambda + \lambda^2 \, \beta_0) \,  {\bf V}_0 \,,
\end{equation}
where
\begin{equation}
\label{Vi operators}
{\bf V}_n | C_{I_0} , C_{I_1},\cdots C_{I_{\ell}}\rangle := \delta^3_{I_n} \, | C_{I_0} , C_{I_1},\cdots C_{I_{\ell}}\rangle \,. \end{equation}
Diagrams contributing to $\beta_0$ are more involved, as some of them include integrals over gluon vertices. Their computation is beyond the scope of our analysis, as we are only interested in the reflection factor at the leading weak-coupling order.

Collecting the expressions \eqref{bulk hamiltonian} and \eqref{bdry hamiltonian} we arrive at
\begin{equation}
\label{semi infinite chain hamitonian}
{\bf H}^{A}= ( \lambda + \lambda^2 \, \beta_0 ) {\bf V}_0 + \lambda^2 \sum_{n=0}^{\infty} ({\bf 1}-{\bf P}_{n,n+1}) \,.
\end{equation}
The above Hamiltonian can be diagonalized by the standard perturbative methods, taking
\begin{equation}
 {\bf H}^0 =  \lambda  {\bf V}_0\,,
 \qquad
 \delta {\bf H} =  \lambda^2 \beta_0  {\bf V}_0
 + \lambda^2 \sum_{n=0}^{\infty} ({\bf 1}-{\bf P}_{n,n+1}) \,.
\end{equation}
Let us consider single impurity states $|n \rangle$, where $n$ indicates the position of the excitation. For these states the unperturbed energies are simply
$E_n^0 = \lambda \delta_{n0}$. However, as the unperturbed spectrum is degenerate, a good basis $| \psi^0_n \rangle$ to apply the methods of perturbation theory should satisfy
\be 
\langle \psi^0_n | \delta H | \psi^0_m \rangle = 0\,,
\qquad\text{for}\quad n\neq m \quad \& \quad 
E^0_n = E^0_m\,.
\label{condi}
\ee 
Thus, we will instead consider the basis
\begin{equation}
|\psi^0_0\rangle := |0\rangle\,, \qquad
|\psi^0(p) \rangle := \sum_{n=1}^{\infty} \left[ e^{-ipn} + R(p) \, e^{ipn} \right] |n \rangle\,, 
\end{equation}
where, in order to satisfy the condition \eqref{condi}, one needs to impose
\begin{equation}
    R(p) = -1\,.
\end{equation}
For a magnon state with momentum $p$ 
the perturbative solution is
\begin{equation}
\label{magnon eigenstates}
|\psi(p) \rangle = \sum_{n=1}^{\infty} \left( e^{-ipn} -  e^{ipn} \right) \, |n \rangle +  \lambda \, (e^{-ip}-e^{ip}) \, |0\rangle + \mathcal{O} (\lambda^2) \,,
\end{equation}
whose energy is
\begin{equation}
\label{magnon energy-perturbative computation}
E(p) = 4 \lambda^2 \sin^2 \frac{p}{2}+ \mathcal{O} (\lambda^3) \,,
\end{equation}
as expected from \eqref{magnon energy}. In addition to the magnon states with momentum $p$ we have a state in which the impurity remains close to the boundary, i.e. a boundary bound state 
\begin{equation}
    \label{boundary bound-states}
    |{\rm B} \rangle = |0 \rangle - \lambda  |1 \rangle + \mathcal{O} (\lambda^2) \,,
\end{equation}
whose energy is
\begin{equation}
\label{bound state energy-perturbative computation}
E_{\rm B} (p, \lambda) = \lambda +  (1+ \beta_0) \, \lambda^2 + \mathcal{O} (\lambda^3) \,.
\end{equation}
One could compute the coefficient $\beta_0$ to know the two-loop correction to the boundary bound-state energy given in \eqref{bound state energy-perturbative computation}, but such computation is not needed to determine the leading weak-coupling correction to the type $A$ magnon dressing phase. Therefore, we conclude that
\begin{equation}
\label{perturbative dressing A}
R_{A}^0(p)=-1 + \mathcal{O} (\lambda^2) \,.
\end{equation}

Finally, let us turn to a weak-coupling analysis 
of the other dressing factor $R_B^0$. We will now consider a $SU(2)$ sub-sector for the even sites of the chain, where the type $B$ impurities can propagate. Therefore, we will work with the states
\begin{equation}
\label{B impurity states}
|\bar C^{I_0} , \bar C^{I_1},\cdots \bar C^{I_{\ell}}\rangle := \sqrt{2} \left( \frac{2k}{N} \right)^{\ell+\frac{1}{2}} \left( \begin{array}{cc} 0 & C_1\bar{C}^{I_0}C_1\bar{C}^{I_1}\cdots \bar{C}^{I_{\ell}}C_{1}
\\ 
0 & 0 \end{array} \right) , 
\end{equation}
where the $\bar C^{I_n}$ are taken to be either $\bar C^{2}$ or $\bar C^{3}$.

A quick diagrammatic analysis shows that the first non-trivial terms in the Hamiltonian ${\bf H}^{B}$ appear at two-loop order. Diagrams that could potentially contribute to a one-loop order, like the ones depicted in Fig. \ref{calzonfermion}, cancel to each other as we have a $C_1$ field at the left-most site. At two-loop order the boundary term in the Hamiltonian acts as the identity operator or it comes with a matrix $M^{I_0}_{J_0}$. 
As the latter does not distinguish between  $\bar C^{2}$ and $\bar C^{3}$, and since we know that $|\bar C^{2} , \bar C^{2},\cdots \bar C^{I_{2}}\rangle$ has vanishing anomalous dimension, the Hamiltonian in this case must be
\begin{equation}
\label{type B hamiltonian}
{\bf H}^{B}= \lambda^2 \sum_{n=0}^{\infty} ({\bf 1}-{\bf P}_{n,n+1})\,.
\end{equation}
For a single magnon impurity we can diagonalize this Hamiltonian with a usual Bethe Ansatz wave-function of the form
\begin{equation}
    |\psi(p) \rangle = \sum_{n=0}^{\infty} \left[ e^{-ipn} + R(p) \, e^{ipn} \right] |n \rangle\,,
\end{equation}
if we fix $R(p)= e^{ip}$. Therefore, we conclude that the type $B$ right-boundary dressing phase is, in the weak coupling limit,
\begin{equation}
\label{weak coupling type B dressing}
R_{B}^0(p)= e^{-ip} + \mathcal{O}(\lambda^2)\,.
\end{equation}

\subsection{All-loop proposal}

We will now make an all-loop proposal for the boundary dressing phases, such that they simultaneously solve the crossing equation discussed in Section \ref{sec: crossing eq} and reproduce the results of the last section when considered in the weak-coupling limit. 

As we have seen, there exists a type $A$ excited state with energy of order $\lambda$. The fact that the dispersion relation \eqref{magnon energy} is expressed as an expansion in even powers of $\lambda$ indicates that such state is not an ordinary magnon, but rather it is a boundary bound state. Let us consider the factor
\begin{equation}
\label{crossing eq factor}
\frac{\frac{1}{x^+}+x^+}{\frac{1}{x^-}+x^-} \,,
\end{equation}
which appears in the r.h.s. of the crossing equation \eqref{crossing eq-2}. Interestingly, it has precisely a pole for $x^- \to -i$ whose energy is\footnote{In all the weak coupling expansions we are using that $h(\lambda) = \lambda+{\cal O}(\lambda^3)$.}
\begin{equation}
\label{energy pole}
E_{\rm pole} = \lambda + \mathcal{O} (\lambda^2) \,,
\end{equation}
in accordance with \eqref{bound state energy-perturbative computation}.  This suggests that the boundary bound state \eqref{boundary bound-states} could arise from a pole if the factor \eqref{crossing eq factor} is included in $R_{A}^0$. Moreover, such factor should be absent in $R_{B}^0$, as we have not  observed boundary bound states associated to type $B$ particles. It is also useful to note that, for real fixed momentum,
\begin{equation}
\frac{\frac{1}{x^+}+x^+}{\frac{1}{x^-}+x^-}
= e^{ip} + {\cal O}(\lambda^2)\,,
\end{equation}
which could serve to explain the relative factor between \eqref{perturbative dressing A} and \eqref{weak coupling type B dressing}.

Taking all these considerations into account, we propose that the all-loop dressing factors are
\begin{equation}
\label{crossing ansatz}
\begin{aligned}
R_{A}^0(p) &= - \frac{1}{R_0 (p)} \left( \frac{\frac{1}{x^+}+x^+}{\frac{1}{x^-}+x^-} \right) \left( \frac{x^-}{x^+} \right)\,, \\
R_{B}^0(p) &= \frac{1}{R_0 (p)} \, \left( \frac{x^-}{x^+} \right)\,,  \\
\end{aligned}
\end{equation}
with
\begin{equation}
\label{crossing eq R0}
R_{0} (p) R_{0} \left( \bar{p} \right) =  \sigma \left( p, -\bar{p} \right) \,,
\end{equation}
and
\begin{equation}
\label{unitarity R0}
R_{0}(-p) R_{0}(p) =1 \,.
\end{equation}
Equations \eqref{crossing eq R0} and \eqref{unitarity R0} can be solved if we take $R_0(p)$ to be the square root of the dressing phase proposed by \cite{Correa:2012hh,Drukker:2012de} for the ${\cal N}=4$ sYM case,
 replacing $g=\frac{\sqrt{\lambda_{YM}}}{4\pi}$ by $h(\lambda)$,
i.e.
\begin{equation}
\label{solution R0}
R_0 (p) = \left[ \frac{1}{\sigma_B (p) \sigma (p,-p)} \left( \frac{1+\frac{1}{(x^-)^2}}{1+\frac{1}{(x^+)^2}} \right) \right]^{\frac{1}{2}} \,,
\end{equation}
with
\begin{equation}
\label{sigma B}
\begin{aligned}
\sigma_B (p) &= e^{i \chi(x^+)-i \chi(x^-)} \,, \\
i \chi (x) &= \left \{ \begin{array}{ccc}
 i \Phi(x)= \oint_{|z|=1} \frac{dz}{2\pi i} \frac{1}{x-z} \log \left\{ \frac{\sinh[2\pi h(z + \frac{1}{z})]}{2\pi h \left( z+\frac{1}{z} \right) } \right\} & {\rm if} & |x|>1 \\
 i \Phi(x)+ \log \left\{ \frac{\sinh[2\pi h(x + \frac{1}{x})]}{2\pi h \left( x+\frac{1}{x} \right) } \right\} & {\rm if} & |x|<1 
\end{array} \right.
\end{aligned}
\end{equation}
Using the results of \cite{Correa:2012hh,Drukker:2012de} we have
\begin{equation}
\label{boundary sigma expansion}
R_0(p)=1+\mathcal{O}(\lambda^2) \,,
\end{equation}
and therefore we recover the expressions \eqref{perturbative dressing A} and \eqref{weak coupling type B dressing} for the weak coupling expansions of the dressing phases $R_{A}^0$ and $R_{B}^0$. Let us note that in order to reproduce \eqref{perturbative dressing A} and \eqref{weak coupling type B dressing} from \eqref{crossing ansatz} we have chosen $r(p)=-1$. 

The fact that $R_0(p)$ is the square root of the dressing factor proposed for the ${\cal N}=4$ sYM case can be naturally understood as follows. In the strong-coupling limit, dressing phases are computed from the scattering of excitations propagating on the worldsheet of strings carrying large angular momentum. These strings propagate in an $AdS_2\times S^2$ sub-space of the geometry (see appendix \ref{app: String theory description}) and, after a Pohlmeyer reduction, the propagation of excitations in the worldsheet is described in terms of sine/sinh Gordon solitons \cite{Hofman:2006xt}. Being the open string restricted to an $AdS_2\times S^2$ sub-space of the geometry, its classical dynamics is identical to that of the open string propagating in $AdS_2\times S^2 \subset AdS_5\times S^5$ \cite{Drukker:2006xg}. Thus, 
the reflection of worldsheet excitations is described in exactly the same way as done in \cite{Correa:2012hh,Drukker:2012de}
\begin{equation}
 R_A^0(p) \simeq R_B^0(p) 
 \simeq e^{-i2T\cos\tfrac{p}{2}\left[\log\left(\frac{1-\sin\frac{p}2}{1+\sin\frac{p}2}\right)+2\log\cos\tfrac{p}{2}\right]}\,,
 \label{stronR0}
\end{equation}
where $T$ is the effective string tension. What changes between one case and the other is how $T$ is related to the 't Hooft coupling. In the  $AdS_4\times {\mathbb{CP}^3}$ background
$2T = \sqrt{2\lambda} \simeq 2 h(\lambda)$ for large $\lambda$. 
Thus, the result \eqref{stronR0}
is in agreement with the strong-coupling limit of our proposal \eqref{crossing ansatz}. Also note that \eqref{stronR0} gives the reflection phase for the two types of magnons. The relative factor in our proposal is an order 1 quantity in the strong-coupling limit and therefore it is not observed in the semiclassical computation.

As said before, the result \eqref{stronR0} also holds for the $AdS_5\times S^5$ background.  However, the relation between the effective string tension and 't Hooft coupling is  $2T = \frac{\sqrt{\lambda_{YM}}}{\pi} = 4 g$ in this case, which explains that the same boundary dressing phase appears squared in the ${\cal N}=4$ sYM case. The fact that bulk S-matrices come with $\sigma(x_1,x_2)$ in one case and with $\sigma(x_1,x_2)^2$ in the other is also explained by the same argument.

Let us note
that type $A$ and type $B$ magnons reflect \textit{differently} at the boundaries. This implies a striking contrast to what is observed for the bulk scattering properties. At weak coupling one can see this as a straightforward consequence of the fact that type $A$ particles are in closer interaction with the boundaries than type $B$ particles. The range of the interactions between the impurities and the boundary is of order $\lambda$ for type $A$ magnons and of order $\lambda^2$ for type $B$ magnons.  As an important implication of this, one type of particle can form a bound state with the boundary while the other can not. 
Let us also mention that this is a distinctive property of the open boundary set by Wilson loops. Extending the ideas of \cite{Berenstein:2005vf}, open spin chains for the anomalous dimensions of determinant operators in ABJM were studied in \cite{Chen:2018sbp,Chen:2019igg}, and no distinction was observed for the reflection of type $A$ and $B$ magnons in those cases.

To conclude this section, let us comment on the possibility of inserting operators with a non-trivial lower off-diagonal block. Alternating $\bar C^2$ with $C_1$ in the lower off-diagonal block does not constitute a supersymmetric insertion. Instead, a possible supersymmetric lower off-diagonal block vacuum, over which magnon excitations can propagate, is the hermitian conjugate of the upper off-diagonal block
\be 
{\cal V}^{\dagger}_\ell = 
\left(
\begin{array}{cc}
 0 & 0 
\\
(\bar C^1 C_2)^\ell \bar C^1 &  0
\end{array} \right)\,.
\label{susyinsertion0 conjugate}
\ee
Thus, the spin-chain vacuum states ${\cal V}_\ell$ and ${\cal V}^{\dagger}_\ell$ are interchanged under charge conjugation.
Consider for example the $U(1)_{\rm extra}$ symmetry of the ABJM spin chain\cite{Klose:2010ki}, under which the fields $\bar{C}^1,\bar{C}^2,C_3,C_4$ have charge +1 and the fields $C_1,C_2,\bar{C}^3$ and $\bar{C}^4$ have charge -1: while ${\cal V}_\ell$ has charge $-1$, ${\cal V}^{\dagger}_\ell$ has charge $+1$. Similarly, type $A$ magnons in the lower off-diagonal block are the charge conjugates of type $B$ magnons in the upper off-diagonal block, and vice-versa. When considering their reflection from the boundaries, type $A$ and $B$ magnons in the lower block behave as type $B$ and $A$ impurities in the upper block, respectively. Therefore, their dressing phases are interchanged. This is manifest in the weak-coupling regime, as type $B$ magnons in the lower block 
interact with the boundaries at order $\lambda$, while the range of the  interaction with the boundaries is of order $\lambda^2$ for type $A$ magnons in the same block.

\section{Cusp anomalous dimension from a set of BTBA equations}
\label{sec: Gamma cusp}

As pointed out in the Introduction, the main goal of this paper is to compute the cusp anomalous dimension $\Gamma_{\rm cusp}$ of ABJM using an integrability approach. In the previous sections we have studied the spin-chain description of the anomalous dimensions of operators inserted at the 1/2 BPS Wilson line of ABJM. In order to compute $\Gamma_{\rm cusp}$ we shall consider instead a cusped Wilson loop, which is correspondingly described by an open spin chain with an appropriate twist in one of its boundaries. We will consider the insertion of the vacuum ${\cal V}_{\ell}$ at the position of the cusp, and we will study the corresponding anomalous dimension as a function of $\ell$. This anomalous dimension is due to finite-size effects, which are included as corrections to the Asymptotic Bethe Ansatz (ABA). Eventually $\Gamma_{\rm cusp}$ would be obtained in the limit in which no operator is inserted at the cusp. The leading finite-size corrections are taken into account by the so called \textit{L\"uscher corrections} and, ultimately, the exact solution can be obtained by making a \textit{Boundary Thermodynamic Bethe Ansatz} (BTBA). The integral BTBA equations that give the finite-size corrections usually can be rewritten as a set of functional equations, known as $Y$-system. In this section we will propose a set of $Y$-system equations for the cusped Wilson loop of ABJM, which in turn we will use to compute the  one-loop cusp anomalous dimension from a leading-order finite-size correction.

\subsection{Y-system for the cusped Wilson line of ABJM}
\label{Ysistem equations}

Given a 1+1-dimensional system of size $L$ at temperature $1/R$, one can obtain its vacuum energy $E_0 (L)$ from the partition function $Z(L,R)$.
More precisely, in the large $R$ limit one has
\begin{equation}
\label{E0 from partition function}
E_0 (L) \sim -\frac{\log Z(L,R)}{R}\,.
\end{equation}
The key to compute $E_0 (L)$ is to make a double Wick rotation 
of the system \cite{Zamolodchikov:1989cf}.
This maps the \textit{physical theory} to a \textit{mirror theory} 
whose energy $\varepsilon$ and momenta $q$ are related to the physical values $E$ and $p$ 
by
\begin{equation}
    \label{double wick rotation}
    p \to i \varepsilon \,, \qquad \text{and} \qquad E \to iq\,.
\end{equation}
This transformation takes the original system with finite volume $L$ and low temperature $1/R\ll 1$ into a system with large volume $R\gg1$ and finite temperature $1/L$. Crucially, for integrable systems, the latter can be studied using ABA tools. In cases with periodic boundary conditions one can use \eqref{E0 from partition function} to compute $E_0(L)$ from the free-energy of the mirror theory. Alternatively, for open boundary conditions the vacuum energy $E_0(L)$ is obtained from the transition amplitude between two boundary states \cite{LeClair:1995uf}. In either case one arrives at 
\begin{equation}
\label{general TBA}
E_0(L)-E_0(\infty)= - \frac{1}{2\pi} \sum_{A} \int_0^{\infty} dq \, \log[1+Y_A(q)]\,,
\end{equation}
where the $Y_A$ functions are the solutions to a set of integral \textit{(B)TBA equations}. 

The BTBA equations can usually be reformulated as a set of functional equations known as the $Y$-system \cite{Gromov:2009tv,Gromov:2009bc,Gromov:2009at,Bombardelli:2009xz}. Interestingly, there are many examples, in particular within the context of the AdS/CFT correspondence, in which the introduction of integrable boundary conditions in a system modifies the analytical and asymptotic properties of the $Y$-functions without changing the $Y$-system. \cite{Correa:2012hh,Drukker:2012de,Behrend:1995zj,OttoChui:2001xx,Gromov:2010dy,Ahn:2010ws,Ahn:2011xq,vanTongeren:2013gva,Bajnok:2012xc}. This could be related to the fact that either for periodic or open boundary conditions in the physical theory, in the mirror theory one deals with exactly the same system of mirror excitations. We will follow this insight  and we will assume that, as it was the case for the ${\cal N}=4$ sYM Wilson loop \cite{Correa:2012hh,Drukker:2012de}, the same $Y$-system that describes the ABJM spectrum with periodic boundary conditions can be used to describe the spectrum with open boundary conditions set by the ABJM Wilson line.
More specifically, we propose that the $Y$-system of the cusped line of ABJM is \cite{Gromov:2009at,Bombardelli:2009xz,Cavaglia:2013lgg}
\begin{align}
\label{y system ABJM-1}
Y_{a,s}^+Y_{a,s}^- &= \frac{(1+Y_{a,s+1})(1+Y_{a,s-1})}{(1+1/Y_{a+1,s})(1+1/Y_{a-1,s})} \, , \qquad s>1, \, (a,s) \neq (2,2) \,, \\
\label{y system ABJM-2}
Y_{a,1}^+Y_{a,1}^- &= \frac{(1+Y_{a,2})(1+Y_{a,0}^I)(1+Y_{a,0}^{II})}{(1+1/Y_{a+1,1})(1+1/Y_{a-1,1})} \,, \\
\label{y system eq momentum carrying nodes-1}
Y_{a,0}^{\alpha,+}  Y_{a,0}^{\beta,-} &= \frac{(1+Y_{a,1})}{(1+1/Y_{a+1,0}^{\beta})(1+1/Y_{a-1,0}^{\alpha})} \,, \qquad a>1 , \, \alpha \neq \beta \,,  \\
\label{y system eq momentum carrying nodes-2}
Y_{1,0}^{\alpha,+}  Y_{1,0}^{\beta,-} &= \frac{(1+Y_{1,1})}{(1+1/Y_{2,0}^{\beta})} \,, \qquad \qquad \qquad \qquad \quad \alpha \neq \beta \,,
\end{align}
where the set of $Y$-functions is given by
\begin{equation}
\label{abjm y functions}
\begin{aligned}
& Y_{a,0}^{\alpha}\,, \, \, a \geq 1, \, \alpha \in \{ I,II \} \,;  \quad  Y_{a,s}\,, (a,s) \in \mathbb{N}^+ \times \mathbb{N}^+, \, a \leq 1 \, \, \text{or} \, \, s \leq 1 \,; \quad \text{and} \quad  Y_{2,2} \,.
\end{aligned}
\end{equation}
In \eqref{y system ABJM-1}-\eqref{y system eq momentum carrying nodes-2} we are writing the $Y$-functions as functions of the spectral parameter $u$, defined as
\begin{equation}
    \label{spectral parameter}
    x(u)+\frac{1}{x(u)}= \frac{u}{h}\,,
\end{equation}
and we are using the notation $f^{[\pm a]}=f(u \pm i a/2)$. Moreover, considering a new set of functions given by
\begin{equation}
\label{abjm t functions}
\begin{aligned}
T_{a,s}^{\alpha} \,, (a,s) \in \mathbb{N} \times \{ -1,0 \}, \,  \alpha \in \{ I, II \} \,; \qquad  T_{a,s}\,, (a,s) \in \mathbb{N} \times \mathbb{N}^+, \, a \leq 2 \, \, \text{or} \, \, s \leq 2 \,,\\
\end{aligned}
\end{equation}
and making the change of variables 
\begin{align}
\label{T to Y-1}
Y_{a,s} &= \frac{T_{a,s+1}T_{a,s-1}}{T_{a+1,s}T_{a-1,s}}\,, \, \quad s \geq 2 \, \qquad a \geq 1 \,, \\
\label{T to Y-2}
Y_{a,1} &= \frac{T_{a,2}T_{a,0}^I T_{a,0}^{II}}{T_{a+1,1}T_{a-1,1}} \,, \, \quad a \geq 1 \,, \\
\label{abjm momentum carrying nodes from t functions}
Y_{a,0}^{\alpha} &= \frac{T_{a,1}T_{a,-1}^{\beta}}{T_{a+1,0}^{\alpha}T_{a-1,0}^{\beta}} \,, \, \quad a \geq 1 \, , \quad \alpha, \beta \in \{ I, II \} \, , \quad \alpha \neq \beta \,,
\end{align}
one can rewrite the $Y$-system equations \eqref{y system ABJM-1}-\eqref{y system eq momentum carrying nodes-2} in terms of a set of Hirota equations\cite{Cavaglia:2013lgg}, 
\begin{align}
\label{T system-1}
T^+_{a,s} T^-_{a,s} &= (1-\delta_{a,0}) \, T_{a+1,s}T_{a-1,s}+T_{a,s+1}T_{a,s-1}\, , \quad s \geq 2 \,,\\
T^+_{a,1} T^-_{a,1} &= (1-\delta_{a,0}) \, T_{a+1,1}T_{a-1,1}+T_{a,2} T_{a,0}^I T_{a,0}^{II} \,, \\
T_{a,0}^{\alpha,+}  T_{a,0}^{\beta,-} &= (1-\delta_{a,0}) \, T_{a+1,0}^{\beta} T_{a-1,0}^{\alpha} +T_{a,-1}^{\alpha} T_{a,1} \, , \quad \alpha, \beta \in \{ I, II \} \, , \quad \alpha \neq \beta \,,\\
\label{T system-4}
 T_{a,-1}^{\alpha,+}   T_{a,-1}^{\beta,-}  &=  T_{a+1,-1}^{\beta} T_{a-1,-1}^{\alpha} \, , \quad a \neq 0 \, , \quad \alpha, \beta \in \{ I, II \} \, , \quad \alpha \neq \beta \,.
\end{align}
As for the formula \eqref{general TBA}, in the ABJM theory it becomes\footnote{Let us note the extra 1/2 factor in the r.h.s. of this equation. This can be explained by taking into account that the dispersion relation of magnons in the ABJM picture has an overall 1/2 factor with respect to the similar dispersion relation of ${\cal N}=4$ sYM \cite{Gromov:2009at}.}
\begin{equation}
\label{TBA ABJM}
E_0(L)-E_0(\infty)= - \frac{1}{4\pi} \sum_{a=1}^{\infty} \int_0^{\infty} dq \, \log[1+Y^I_{a,0}(q)]- \frac{1}{4\pi} \sum_{a=1}^{\infty} \int_0^{\infty} dq \, \log[1+Y^{II}_{a,0}(q)] \,.
\end{equation}

\subsection{Asymptotic solution of the $Y$-system}
\label{asymptotic solution}

In order to obtain the leading-order contribution to $\Gamma_{\rm cusp}$ from \eqref{TBA ABJM} we will discuss now the asymptotic large-volume solution that is obtained from the $Y$-system of eqs. \eqref{y system ABJM-1}-\eqref{y system eq momentum carrying nodes-2}. Following \cite{Gromov:2009at}, in the asymptotic limit one gets
\begin{align}
\label{asymptotic solution 1}
Y^I_{a,0} &\sim \left( \frac{z^{[-a]}}{z^{[+a]}} \right)^{2L} \, T_{a,1} \, \prod_{n=-\frac{a-1}{2}}^{\frac{a-1}{2}} \phi_{I}^{\zeta(n,a)} \left( u + i n \right) \phi_{II}^{1-\zeta(n,a)} \left( u + i n \right) \,,\\
\label{asymptotic solution 2}
Y^{II}_{a,0} &\sim \left( \frac{z^{[-a]}}{z^{[+a]}} \right)^{2L} \, T_{a,1} \, \prod_{n=-\frac{a-1}{2}}^{\frac{a-1}{2}} \phi_{II}^{\zeta(n,a)} \left( u + i n \right) \phi_{I}^{1-\zeta(n,a)} \left( u + i n \right) \,,
\end{align}
where the two functions $\phi_I$ and $\phi_{II}$ will be fixed later by comparison with L\"uscher corrections, and with
\begin{equation*}
\label{zeta}
\zeta(n,a) = \left\{ 
\begin{array}{cc}
1     & \text{if }  n+\frac{a-1}{2} \text{ is even} \\
0     & \text{if }  n+\frac{a-1}{2} \text{ is odd}
\end{array} \right.
\end{equation*}
In \eqref{asymptotic solution 1} and \eqref{asymptotic solution 2} we are using the notation $z^{\pm}$ for the spectral variables in the mirror theory. In the particular case in which 
\begin{equation}
\label{phi equality0}
\phi_I(u)=\phi_{II}(u)= \pm \frac{\varphi \left( u- \frac{i}{2} \right)}{\varphi \left( u+ \frac{i}{2} \right)} \,,
\end{equation}
for some function $\varphi$, one gets
\begin{equation}
\label{phi equality}
Y^I_{a,0}=Y^{II}_{a,0} \sim (\pm 1)^a \left( \frac{z^{[-a]}}{z^{[+a]}} \right)^{2L} \, \frac{\varphi \left( u- \frac{ia}{2} \right)}{\varphi \left( u+ \frac{ia}{2} \right)} \, T_{a,1} \,.
\end{equation}

Let us start by discussing the $T_{a,1}$ functions that appear in \eqref{asymptotic solution 1} and \eqref{asymptotic solution 2}, which can be obtained as a solution of the $T$-system of eqs. \eqref{T system-1}-\eqref{T system-4}.  In principle, one could determine the asymptotic $T_{a,1}$ functions with a computation of the double-row transfer matrix for a bound state of $a$ magnons. Instead of directly computing such a double-row transfer matrix we will follow the ideas of \cite{Bajnok:2012xc}, and we will argue that the $T_{a,1}$ functions of a system with $SU(1|2)$ symmetry can be obtained from the corresponding $T$-functions of a system with $SU(2|1)$ symmetry from the identification
\begin{equation}
T_{a,1}^{SU(1|2)} \equiv \left( T_{1,a}^{SU(2|1)} \right)^*\,.
\end{equation}
The $T$-system of the case with $SU(2|1)$ symmetry has already been studied in \cite{Bajnok:2013wsa}. Following their results, we obtain
\begin{equation}
\label{asymptotic T functions}
T_{a,1}^{SU(1|2)}= 2 (-1)^a \left[ b_{0,a} \left( 1+ \frac{u^{[-a]}}{u^{[a]}} \right)  + 2 \sum_{k=1}^{a-1} \frac{b_{k,a} u^{[-a]}}{u^{[-a+2k]}} \right] \, , a \geq 1  \,, \\
\end{equation}
with
$$
\begin{aligned}
b_{0,s}&= \sin^2 \tfrac{\theta}{2} \, P_{s-1}^{(0,1)} \left( 1-2\cos^2 \tfrac{\theta}{2} \right) \,, \\
b_{l,s}&=b_{s-l,s}=b_{0,l}b_{0,s-l} \,,\\
b_{0,0}&=1 \,,
\end{aligned}
$$
and where $P_{s-1}^{(0,1)}$ stands for Jacobi polynomial. For simplicity, from now on we will simply write $T_{a,1}^{SU(1|2)} \equiv T_{a,1}$. The asymptotic solution to the $T$-system proposed in \eqref{T system-1}-\eqref{T system-4} is completed by\footnote{This solution receives finite-size corrections away from the asymptotic limit. In particular, following \eqref{abjm momentum carrying nodes from t functions} we see that the correction to $T_{a,-1}^{\alpha}$ with $\alpha=I,II$ and $a \geq 1$ gives the leading-order contribution to $Y_{a,0}^{\alpha}$, which is presented in \eqref{asymptotic solution 1} and \eqref{asymptotic solution 2}.}
\begin{equation}
\label{cusped WL t system ABJM}
\begin{aligned}
T_{1,s}	&= (-1)^s \frac{4su}{u^{[s]}} \sin^2 \frac{\theta}{2} \, , \, \, s \geq 1 \,; \qquad 
T_{2,s}=T_{s,2}= \frac{16 u^{[s]} u^{[-s]}}{u^{[s-1]u^{[s+1]}}} \sin^4 \frac{\theta}{2} \, , \, \, s \geq 2\,; \\
T_{0,s} &= 1 \, , \, \, s > 0 \,; \qquad \qquad \quad \qquad \quad \; \,
T_{a,0}^I =1 \, , \, \, a \geq 0 \,; \\
T_{a,0}^{II} &=1 \, , \, \, a \geq 0 \,; \qquad \qquad \quad \qquad \; \; \;
T_{0,-1}^I =1 \,;  \\
T_{0,-1}^{II} &=1 \,; \qquad \qquad \quad \qquad \quad \qquad \quad \,
T_{a,-1}^I =0 \, , \, \, a \geq 1 \,; \\
T_{a,-1}^{II} &=0 \, , \, \, a \geq 1 \,.
\end{aligned}
\end{equation}

Having discussed the asymptotic $T_{a,1}$ functions associated with the cusped Wilson line, let us focus now on the $\phi_I$ and $\phi_{II}$ functions that appear in the asymptotic solutions \eqref{asymptotic solution 1} and \eqref{asymptotic solution 2}. 
We will obtain an expression for these functions from the study of the leading-order L\"uscher correction to the vacuum energy $E_0$,
which for the ABJM cusped Wilson line we propose to be given as
\begin{align}
\label{Luscher ABJM-1}
Y_{a,0}^I (q) &\sim e^{-2L \, \epsilon_a(q)} \, {\rm Tr} \left[ R_{A}^{\it up}(q) \, C \, R_{B, \theta}^{\it down} \, (-\bar{q}) \, C^{-1} \right] \,, \\
\label{Luscher ABJM-2}
Y_{a,0}^{II} (q) &\sim e^{-2L \, \epsilon_a(q)} \, {\rm Tr} \left[ R_{B}^{\it up}(q) \, C \, R_{A,\theta}^{\it down} \, (-\bar{q}) \, C^{-1} \right] \,.
\end{align}
Let us comment some details about the above formulas. First, the function $\epsilon_a(q)$ gives the energy of an $a$-magnon of mirror-momentum $q$, and
\begin{equation}
\label{charge conjugation}
C= \left( \begin{array}{cc}
-i \epsilon_{ab} & 0 \\
0 & \epsilon_{\alpha \beta}
\end{array}
\right),
\end{equation}
is the charge-conjugation matrix. Moreover, $R_{A/B}$ is the reflection matrix derived in the previous sections for a Wilson line along the $x_0$ direction, while $R_{A/B,\theta}$ is the corresponding reflection matrix for a Wilson loop rotated by an angle $\theta$, 
\begin{equation}
\label{rotated reflection}
R_{A/B,\theta} (q) = {\cal O} (\theta) R_{A/B} (q) {\cal O}^{-1} (\theta) \,,
\end{equation}
where the rotation matrix ${\cal O} (\theta)$ is given by
\begin{equation}
\label{rotation matrix}
{\cal O} (\theta)= \left( 
\begin{array}{cccc}
1 & 0 & 0 & 0 \\
0 & 1 & 0 & 0 \\
0 & 0 & \cos \frac{\theta}{2} & \sin \frac{\theta}{2} \\
0 & 0 & -\sin \frac{\theta}{2} & \cos \frac{\theta}{2} \\
\end{array}
\right) \,.
\end{equation}
Note that we are only considering the case of a Wilson line with a geometric cusp $\theta$, i.e. we are not including an internal cusp angle $\varphi$\footnote{The internal cusp $\varphi$ should account for an R-symmetry rotation in a plane that includes the $I=1$ direction. However, such a rotated line would not share any supersymmetries with the vacuum state written in \eqref{susyinsertion0}, which would notoriously difficult the computation of the corresponding rotation matrix.}. The words ${\it up}$ and ${\it down}$
in \eqref{Luscher ABJM-1} and \eqref{Luscher ABJM-2}
refer to the reflection matrices of impurities in the upper or lower off-diagonal blocks
of the corresponding supermatrix, respectively.
It is crucial to 
recall that the charge conjugate of a type $A$ (or $B$) magnon 
in the upper off-diagonal block spin chain is a type $B$ (or $A$) magnon in the lower off-diagonal block spin chain. Then, taking into account that
\begin{equation}
\begin{aligned}
R_{A}^{\it down}(q) &= R_{B}^{\it up}(q) \,, \\
R_{B}^{\it down}(q) &= R_{A}^{\it up}(q) \,, \\
\end{aligned}
\end{equation}
we arrive at
\begin{align}
\label{Luscher ABJM-3}
Y_{a,0}^I (q) &\sim e^{-2L \, \epsilon_a(q)} \, {\rm Tr} \left[ R_{A}^{\it up}(q) \, C \, R_{A, \theta}^{\it up} \, (-\bar{q}) \, C^{-1} \right] \,, \\
\label{Luscher ABJM-4}
Y_{a,0}^{II} (q) &\sim e^{-2L \, \epsilon_a(q)} \, {\rm Tr} \left[ R_{B}^{\it up}(q) \, C \, R_{B,\theta}^{\it up} \, (-\bar{q}) \, C^{-1} \right] \,.
\end{align}
Otherwise stated, from now on we will return to work only with reflection matrices of magnons in the upper off-diagonal block. Moreover, to simplify the notation we will again refer to them simply as $R_{A}$ and $R_{B}$.

Let us turn now to the explicit computation of \eqref{Luscher ABJM-3} and \eqref{Luscher ABJM-4}, and let us focus first on the computation of $Y_{1,0}^I$ and $Y_{1,0}^{II}$. Using the results of Sections \ref{sec: open spin chain} and \ref{sec: crossing} we arrive at
\begin{align}
\label{rotated transfer matrix-trace}
Y_{1,0}^I(z^+,z^-) &= e^{-2L \, \epsilon_1} \, \sin^2 \frac{\theta}{2}\, \frac{(z^+ + z^-)^2}{z^+ z^-} \, R_{A}^0 \left( z^+ , z^- \right) R_{A}^0 \left( -\tfrac{1}{z^-} , -\tfrac{1}{z^+} \right) \,, \\
\label{rotated transfer matrix-trace-2}
Y_{1,0}^{II}(z^+,z^-) &= e^{-2L \, \epsilon_1} \, \sin^2 \frac{\theta}{2} \, \frac{(z^+ + z^-)^2}{z^+ z^-} \, R_{B}^0 \left( z^+ , z^- \right) R_{B}^0 \left( -\tfrac{1}{z^-} , -\tfrac{1}{z^+} \right) \,.
\end{align}
Moreover, from \eqref{crossing ansatz} we get
\begin{equation}
\label{a magnon-a=1-2}
\begin{aligned}
R_{A}^0 \left( z^+ , z^- \right) R_{A}^0 \left(- \tfrac{1}{z^-} , -\tfrac{1}{z^+} \right)=R_{B}^0 \left( z^+ , z^- \right) R_{B}^0 \left(- \tfrac{1}{z^-} , -\tfrac{1}{z^+} \right) \,, \\
\end{aligned}
\end{equation}
which implies
\begin{equation}
\label{a magnon-a=1-3}
\begin{aligned}
Y_{1,0}^I(z^+,z^-)=Y_{1,0}^{II}(z^+,z^-)= e^{-2L \, \epsilon_1} \, \sin^2 \frac{\theta}{2} \, \frac{(z^+ + z^-)^2}{z^+ z^-} \, R_{A}^0 \left( z^+ , z^- \right) R_{A}^0 \left( -\tfrac{1}{z^-} , -\tfrac{1}{z^+} \right) \,.
\end{aligned}
\end{equation}
Suggestively, one can rewrite \eqref{rotated transfer matrix-trace} and \eqref{rotated transfer matrix-trace-2} as
\begin{equation}
\label{rotated transfer matrix-trace-a1-1}
Y_{1,0}^I= Y_{1,0}^{II}= - e^{-2L \, \epsilon_1} \frac{(z^+ + z^-)^2}{2 \, z^+ z^- \left( 1+\frac{z^-+\frac{1}{z^-}}{z^+ +\frac{1}{z^+}}\right)}  \, R_{A}^0 \left( z^+ , z^- \right) R_{A}^0 \left( -\tfrac{1}{z^-} , -\tfrac{1}{z^+} \right) \, T_{1,1} \,,
\end{equation}
where $T_{1,1}$ is given in \eqref{asymptotic T functions}. 

Let us focus for the moment on the product of dressing phases that appears in \eqref{rotated transfer matrix-trace-a1-1}. Using \eqref{crossing ansatz} and \eqref{solution R0} we have
\begin{equation}
\label{product dressing phases}
\begin{aligned}
R_{A}^0 \left( z^+ , z^- \right) R_{A}^0 \left( -\tfrac{1}{z^-} , -\tfrac{1}{z^+} \right)  &= \sigma_B^{1/2}(q) \, \sigma_B^{1/2}(-\bar{q}) \, \sigma^{1/2} (q,-q) \, \sigma^{1/2} (-\bar{q},\bar{q}) \, \left( \frac{z^-}{z^+} \right)^3 \,.
\end{aligned}
\end{equation}
The product of bulk dressing phases that appear in \eqref{product dressing phases} can be evaluated using the identity \cite{Correa:2009mz}
\begin{equation}
\sigma(q_i,-q_j) \sigma(-\bar{q}_i,\bar{q}_j)=  \frac{z_i^+}{z_i^-} \frac{z_j^+}{z_j^-}  \frac{f(q_j,-q_i)}{f(q_j,-\bar{q}_i)} \,,    
\end{equation}
with
\begin{equation}
f(z_1,z_2):=\frac{z_1^- -z_2^+}{z_1^- -z_2^-} \frac{1-1/z_1^+ z_2^+}{1-1/z_1^+ z_2^-} \,,
\end{equation}
which gives
\begin{equation}
\sigma^{1/2}(q,-q) \sigma^{1/2}(-\bar{q},\bar{q})= 2 \,  \frac{z^+}{z^-} \left( \frac{1+z^+ z^-}{z^+ + z^-} \right)\frac{1}{\sqrt{\left(z^+ + \frac{1}{z^+} \right) \left(z^- + \frac{1}{z^-} \right)}}  \,.    
\end{equation}
Therefore,
\begin{equation}
\label{product dressing phases-2}
\begin{aligned}
R_{A}^0 \left( z^+ , z^- \right) R_{A}^0 \left( -\tfrac{1}{z^-} , -\tfrac{1}{z^+} \right)  &= 2 \, \sigma_B^{1/2}(q) \, \sigma_B^{1/2}(-\bar{q}) \, \left( \frac{z^-}{z^+} \right)^2 \left( \frac{1+z^+ z^-}{z^+ + z^-} \right)\, \\
& \quad \, \times  \frac{1}{\sqrt{\left(z^+ + \frac{1}{z^+} \right) \left(z^- + \frac{1}{z^-} \right)}}  \,,
\end{aligned}
\end{equation}
and
\begin{equation}
\label{Y1}
Y_{1,0}^I= Y_{1,0}^{II}= - e^{-(2L+2) \, \epsilon_1} \, \sigma_B^{1/2}(q) \, \sigma_B^{1/2}(-\bar{q}) \, \left( \frac{z^+ + \tfrac{1}{z^+}}{z^- + \tfrac{1}{z^-}}\right)^{1/2}   T_{1,1} \,.
\end{equation}
From \eqref{Y1} we can read the expressions for the $\phi_I$ and $\phi_{II}$ functions introduced in \eqref{asymptotic solution 1} and \eqref{asymptotic solution 2}, which we see that behave as in \eqref{phi equality0}. Consequently, following \eqref{phi equality} we can propose
\begin{equation}
\label{Y functions-1}
Y_{a,0}^I= Y_{a,0}^{II}= (-1)^a e^{-(2L+2) \, \epsilon_a} \, \sigma_B^{1/2}(q) \, \sigma_B^{1/2}(-\bar{q}) \, \left( \frac{z^{[+a]} + \tfrac{1}{z^{[+a]}}}{z^{[-a]} + \tfrac{1}{z^{[-a]}}}\right)^{1/2}    T_{a,1} \,,
\end{equation}
for general $a$. Taking into account the discussion given at the beginning of this section, we get that the expressions \eqref{Y functions-1} constitute an asymptotic solution to the Y-system presented in  \eqref{y system ABJM-1}-\eqref{y system eq momentum carrying nodes-2}.

\subsection{Cusp anomalous dimension from the BTBA formula}

Let us focus now on the computation of the leading-order contribution to $E_0(L)-E_0(\infty)$, that comes from inserting the asymptotic solutions given in \eqref{Y functions-1} into the BTBA formula presented in \eqref{TBA ABJM}.
We will compute the finite-size correction at leading weak-coupling order. As we will see next, the explicit evaluation of \eqref{Y functions-1} shows that
\begin{equation}
Y^{I}_{a,0} \sim Y^{II}_{a,0} \sim 
{\cal O}(h^{4L+4})
\,,
\end{equation}
For the leading-order contribution to \eqref{TBA ABJM} there are two distinct possibilities, depending on whether each $Y$-function has a double pole or not as $q\to 0$ \cite{Correa:2009mz}. On the one hand, for $Y$-functions that are regular as $q\to 0$ one gets
\begin{equation}
\log \left( 1+ Y^{\alpha}_{a,0} \right) \sim Y^{\alpha}_{a,0} \sim 
{\cal O}(h^{4L+4}) \,, 
\end{equation}
which specifies the order of the finite-size correction. This leading asymptotic contribution is mediated by the exchange of a two-particle state in the mirror theory. On the other hand, if a $Y$-function has a double pole in $q$, i.e.
\begin{equation}
Y^{\alpha}_{a,0} = \frac{C_{\alpha,a}^2}{q^2} + \mathcal{O} (1) \,,
\end{equation}
the integral of its contribution to the finite-size correction provides a term proportional to
\begin{equation}
\label{approx TBA double pole}
 C_{\alpha,a} 
 \sim  {\cal O}(h^{2L+2}) \,.
\end{equation}
In this case,  the correction is mediated by the exchange of a one-particle state in the mirror picture.

Let us therefore study the behaviour of the $Y$-functions given in \eqref{Y functions-1} for $q \to 0$ and in the leading weak-coupling limit. From the results of \cite{Correa:2012hh,Drukker:2012de}, the factor that includes the boundary phase $\sigma_B$ behaves as
\begin{equation}
\sigma_B^{1/2} (q) \sigma_B^{1/2} (-\bar{q}) = \left(\frac{a}{q} + {\cal O}(q^0)\right) + {\cal O}(h^2)\,.
\label{polosigmaB}
\end{equation}
Furthermore, we have
\begin{equation}
e^{-(2 L+2) \, \epsilon_{a}(q)} = \left( \frac{4h^2}{a^2+q^2} \right)^{2L+2} +
{\cal O}(h^{4L+6})\,.
\label{exponopole}
\end{equation}
 It remains to evaluate the $T$-functions for $q \to 0$ to leading weak-coupling order.  For odd values of $a=2n+1$, the $T_{a,1}$ vanish linearly in $q$. Thus, altogether with the other factors \eqref{polosigmaB} and \eqref{exponopole}, we have regular $Y$-functions in the limit $q\to 0$,
\begin{equation}
\label{approx Y odd}
 Y^{I}_{2n+1,0} = Y^{II}_{2n+1,0} =\mathcal{O}(q^0) \,.
\end{equation}
On the other hand, for even values of $a$ the $T$-functions present a simple pole
\begin{equation}
\label{even a T functions expansion}
\left( \frac{z^{[+2n]} + \tfrac{1}{z^{[+2n]}}}{z^{[-2n]} + \tfrac{1}{z^{[-2n]}}}\right)^{1/2} T_{2n,1} = \left( \frac{8n b_{0,n}^2}{q}  + {\cal O}(q^{0}) \right) + {\cal O}(h^2) \,,
\end{equation}
Therefore, for even values of $a$,
\begin{equation}
\label{approx Y even}
Y^{I}_{2n,0} = Y^{II}_{2n,0} = \left( \frac{16 \, b_{0,n}^2 h^2}{q^2} \left( \frac{h}{n} \right)^{4L+2} + {\cal O}(q^0) \right) + {\cal O}(h^{4L+6}) \,.
\end{equation}
 Plugging these asymptotic expressions in \eqref{Y functions-1} we obtain
\begin{align}
E_0 (L) - E_0 (\infty) &= - 2h^{2L+2} \sum_{n=1}^{\infty}\frac{b_{0,n}}{n^{2L+1}}  + \mathcal{O}(h^{2L+3})\nonumber
\\ 
&=  - 2h^{2L+2} \sin^2\tfrac{\theta}{2} \sum_{k=0}^{\infty}
\frac{P_k^{(0,1)}(-\cos\theta)}{(k+1)^{2L+1}}  + \mathcal{O}(h^{2L+3}) \,.
\label{TBA result-L}
\end{align}

Let us comment about the sign of \eqref{TBA result-L}. The result of the integrals over $q$ that appear in \eqref{TBA ABJM} are given by the square root of the coefficient in front of the double poles in \eqref{approx Y even}, which entails  an ambiguity in the sign choice of the result. The correct sign can be determined if we consider the limit $\theta\to\pi$, at which the Wilson loop configuration can be related to a quark antiquark pair and  each term should contribute negatively to the energy \cite{Correa:2012hh}. The Jacobi polynomials are normalized such that
$P_k^{(0,1)}(1) = 1$, and this fixes the sign  to be the one in \eqref{TBA result-L}.

Eq. \eqref{TBA result-L} gives the finite-size correction to the energy  of a vacuum state \eqref{susyinsertion0} inserted at the position of the cusp. Its evaluation for arbitrary $L$ is rather complicated. If we identify the TBA-length $L$ with $\ell$, 
the vacuum state insertion is made of $2L+1$ fields. In order to associate the finite-size correction to this vacuum energy with $\Gamma_{\rm cusp}$, we need to consider an insertion with a vanishing number of fields, which
requires to analytically continue  the length $L$ such that
\begin{equation}
2L+1=0 
\qquad
\Rightarrow
\qquad
L = -1/2
\,.
\end{equation}
Therefore, we are interested in computing
\begin{equation}
\label{TBA result-2}
E_0 (-1/2) - E_0 (\infty) = - 2h \, \sin^2 \frac{\theta}{2} \sum_{k=0}^{\infty} P_{k}^{(0,1)} (-\cos\theta)  + \mathcal{O}(h^2) \,.
\end{equation}
Using the generating function of Jacobi Polynomials 
\begin{equation}
\label{Jacobi sum identity}
\sum_{k=0}^{\infty} P_{k}^{(0,1)} \left( x \right) t^k = \frac{2}{(1+t+ \sqrt{1-2x t + t^2})\sqrt{1-2x t + t^2}} \,,
\end{equation}
it is straightforward to compute the sum that appears in \eqref{TBA result-2}, which gives
\begin{equation}
\label{TBA result-3}
E_0 (-1/2) - E_0 (\infty) = -  h \left( \frac{1}{\cos \frac{\theta}{2}}-1 \right) + \mathcal{O}(h^2) \,.
\end{equation}

As shown in \cite{Minahan:2009aq,Leoni:2010tb}, in the weak-coupling limit the interpolating function behaves as $
h(\lambda) = \lambda + \mathcal{O}(\lambda^2) $. Consequently,
 the result \eqref{TBA result-3} precisely agrees with the one-loop cusp anomalous dimension \eqref{perturbative light like gamma cusp}, computed in \cite{Griguolo:2012iq} from an expansion in  Feynman diagrams.

\subsection{BTBA equations for the cusped Wilson line of ABJM}
\label{BTB equations}

Finally, let us write the BTBA equations for the cusped Wilson line of ABJM. To that aim, we should recall that in the previous sections we used that the $Y$-system for the cusped Wilson line of ABJM is the same as the one that describes the corresponding periodic system (the only changes are in the analytical and asymptotic properties of the solutions).  The same was observed in the ${\cal N}=4$ sYM, and the BTBA equations for the cusped Wilson line in that case were found to be almost the same as the TBA equations for the periodic spin chain. To be more precise, one can obtain the BTBA equations by taking the TBA equations of the periodic system and then subtracting the result of evaluating them in the leading-order finite-size solution  \cite{Correa:2012hh}. 
Assuming that the same holds in the ABJM case and using the TBA equations presented in \cite{Gromov:2009at,Bombardelli:2009xz} for the periodic spin chain, we propose that the BTBA equations for the cusped Wilson line of ABJM are
\begin{align}
\log \left( \frac{Y_{1,1}}{\mathbf{Y}_{1,1}} \right) &= K_{m-1} \star \log \left( \frac{1+ \bar{Y}_{1,m}}{1+Y_{m,1}} \frac{1+\mathbf{Y}_{m,1}}{1+ \mathbf{\bar{Y}}_{1,m}} \right) + {\cal R}^{(01)}_{1m} \star \log (1+Y_{m,0}^I) + {\cal R}^{(01)}_{1m} \star \log (1+Y_{m,0}^{II}) \,, \nonumber \\
\log \left( \frac{\bar{Y}_{2,2}}{\mathbf{\bar{Y}}_{2,2}} \right) &= K_{m-1} \star \log \left( \frac{1+ \bar{Y}_{1,m}}{1+Y_{m,1}} \frac{1+\mathbf{Y}_{m,1}}{1+ \mathbf{\bar{Y}}_{1,m}} \right) + {\cal B}^{(01)}_{1m} \star \log (1+Y_{m,0}^I) + {\cal B}^{(01)}_{1m} \star \log (1+Y_{m,0}^{II}) \,, \nonumber \\
\log \left( \frac{\bar{Y}_{1,n}}{\mathbf{\bar{Y}}_{1,n}} \right) &= - K_{n-1,m-1} \star \log \left( \frac{1+\bar{Y}_{1,m}}{1+\mathbf{\bar{Y}}_{1,m}} \right) - K_{n-1} \circledast \log \left( \frac{1+Y_{1,1}}{1+\mathbf{Y}_{1,1}} \right) \,, \nonumber \\
\log \left( \frac{Y_{n,1}}{\mathbf{Y}_{n,1}} \right) &= - K_{n-1,m-1} \star \log \left( \frac{1+Y_{m,1}}{1+\mathbf{Y}_{m,1}} \right) -  K_{n-1} \circledast \log \left( \frac{1+Y_{1,1}}{1+\mathbf{Y}_{1,1}} \right)  + \nonumber \\
&\quad + \left( {\cal R}^{(01)}_{nm} +{\cal B}^{(01)}_{n-2,m}  \right) \star \log (1+Y_{m,0}^I) + \left( {\cal R}^{(01)}_{nm} +{\cal B}^{(01)}_{n-2,m}  \right) \star \log (1+Y_{m,0}^{II}) \,, \nonumber \\
\log \left( \frac{Y_{n,0}^I}{\mathbf{Y}_{n,0}^I} \right) &= {\cal T}^{\parallel}_{nm} \star \log (1+ Y_{m,0}^I) + {\cal T}^{\perp}_{nm} \star \log (1+ Y_{m,0}^{II}) \, + \nonumber \\
& \quad + {\cal R}^{(10)}_{n1} \circledast \log \left( \frac{1+Y_{1,1}}{1+\mathbf{Y}_{1,1}} \right) + \left( {\cal R}^{(10)}_{nm} +{\cal B}^{(10)}_{n,m-2}  \right) \star \log \left( \frac{1+Y_{m,1}}{1+\mathbf{Y}_{m,1}} \right) \,, \nonumber 
\end{align}
\begin{align}
\log \left( \frac{Y_{n,0}^{II}}{\mathbf{Y}_{n,0}^{II}} \right) &= {\cal T}^{\parallel}_{nm} \star \log (1+ Y_{m,0}^{II}) + {\cal T}^{\perp}_{nm} \star \log (1+ Y_{m,0}^{I}) \, + \nonumber \\
& \quad + {\cal R}^{(10)}_{n1} \circledast \log \left( \frac{1+Y_{1,1}}{1+\mathbf{Y}_{1,1}} \right) + \left( {\cal R}^{(10)}_{nm} +{\cal B}^{(10)}_{n,m-2}  \right) \star \log \left( \frac{1+Y_{m,1}}{1+\mathbf{Y}_{m,1}} \right) \,. \nonumber
\end{align}
Above we are following the conventions of \cite{Gromov:2009at} for the integral kernels and convolutions, and we are using the notation $\bar{Y}_{a,s}=1/Y_{a,s}$. Moreover, the bold face $\mathbf{Y}$'s are used to represent the leading-order finite-size solutions, whose explicit expressions can be obtained from \eqref{T to Y-1}, \eqref{T to Y-2}, \eqref{asymptotic T functions}, \eqref{cusped WL t system ABJM} and \eqref{Y functions-1}.  Let us note that, working as in \cite{Gromov:2009bc,Bombardelli:2009xz,Gromov:2009at}, one can see that the BTBA equations that we have proposed in this section consistently lead to the $Y$-system written in \eqref{y system ABJM-1}-\eqref{y system eq momentum carrying nodes-2}.

\section{Conclusions}
\label{sec: conclusions}

 In this paper we have proposed a $Y$-system of equations for the cusped Wilson line of the three-dimensional ${\cal N}=6$ Chern-Simons-matter (ABJM) theory, and we have shown that those equations consistently reproduce the one-loop cusp anomalous dimension $\Gamma_{\rm cusp}$ of ABJM from a leading-order finite-size correction. Moreover, we have proposed  a compatible set of Boundary Thermodynamic Bethe Ansatz (BTBA) equations. To perform the BTBA analysis we have constructed an integrable open spin chain that allows to describe the insertion of operators along the contour of the 1/2 BPS Wilson loop of ABJM. Furthermore, we have computed the corresponding reflection matrices,
including  an all-loop proposal for their dressing,
and we have shown the consistence of our proposal with the expected weak- and strong-coupling behaviours.

There are many exciting open questions that arise from our results. In first place, it would be interesting to use the BTBA equations of the cusped Wilson line to derive the bremmstrahlung function of ABJM, as done in \cite{Gromov:2012eu} for the ${\cal N}=4$ super Yang-Mills (sYM) theory. A comparison of this result with the localization-based computation of \cite{Bianchi:2017svd,Bianchi:2018scb} would allow for a direct derivation of the interpolating function $h(\lambda)$ of ABJM, in terms of which every all-loop integrability results are expressed. The result of such computation could be then compared to the current conjecture given in \cite{Gromov:2014eha}.

Finite-size corrections to the vacuum energy of the cusped Wilson line of ${\cal N}=4$ sYM were reformulated in terms of the Quantum Spectral Curve (QSC) formalism in \cite{Gromov:2015dfa}. There the authors found that the functional relations of the QSC were the same as for the corresponding periodic system, and only the asymptotic and analytic properties of the solutions had to be modified. This is suggestively similar to what we have found for the $Y$-system   in our setup. It would be interesting to follow this insight in order to propose a Quantum Spectral Curve for  the cusped Wilson line in ABJM.

Furthermore, it would be interesting to study the generalization of our results to the ABJ theory \cite{Aharony:2008gk}, i.e. to the case in which the ranks of the gauge groups are not necessarily equal. In this picture, evidence of integrability was found in \cite{Minahan:2009aq} in the weak-coupling limit, while an all-loop proposal for the interpolating function $h$ was made in \cite{Cavaglia:2016ide}. However, the all-loop integrability of the ABJ theory is not trivially guaranteed from the assumed integrability of the ABJM limit. It is known that the string sigma model dual to the ABJ theory contains a theta-angle term proportional to $\lambda-\hat{\lambda}$ \cite{Aharony:2008gk}, where $\lambda$ and $\hat{\lambda}$ are the 't Hooft couplings of the ABJ theory. Such term implies a violation of parity, which is sometimes related to a breakdown of integrability (see for example the discussion in \cite{Minahan:2009te}). In this regard, it is suggestive to note that the 
generalization of the boundary Hamiltonian ${\bf H}^A_{\rm bdry}$ discussed in Section \ref{sec: weak coupling} 
would have both an order $\lambda$ and an order $\hat{\lambda}$ term when computed in the ABJ theory,  which would also imply a violation of parity.

One crucial ingredient of our results is the all-loop proposal for the boundary dressing phases. It would be interesting to further check the solutions to the crossing-unitarity equations that we have given, for example by a two-loop computation of the energy of the boundary bound states.

Finally, it would also be interesting to extend the TBA-based computation of $\Gamma_{\rm cusp}$ to the next-to-leading order, as done for the ${\cal N}=4$ sYM case in \cite{Bajnok:2013sya}. This would require an iteration of the BTBA equations, whose  result could eventually be compared with the Feynman-diagram computation done in \cite{Griguolo:2012iq}.
This would constitute a test for the proposed dressing factors as well as for the BTBA equations.

\section*{Acknowledgements}

We would like to thank C. Ahn, Z. Bajnok, J. Balog, L. Bianchi, A. Cavagli\`a, N. Gromov, J. Miczajka and M. Preti for useful discussions. We would like to especially thank J.Aguilera-Damia for collaboration at the early stages of this work for the string theory description of the problem. This work was partially supported by  PICT 2020-03749, PIP 02229, UNLP X910 and PUE084 ``B\'usqueda de nueva f\'isica''.
DHC would like to acknowledge support from the ICTP through the Associates Programme (2020-2025). ML is supported by fellowships from CONICET (Argentina) and DAAD (Germany).

\appendix

\section{ABJM conventions}
\label{app: ABJM conventions}

The ABJM theory is a three-dimensional ${\cal N}=6$ Chern-Simons-matter theory with gauge group $U(N) \times U(N)$. The field content of the theory is given by gauge fields $\left( A_{\mu} \right)^i_j$ and $( \hat{A}_{\mu} )^{\hat{i}}_{\hat{j}}$ in the adjoint representations of the corresponding gauge groups, bi-fundamental complex scalar fields $( C_I )^{\hat{i}}_j$ and $( \bar{C}^I)^{i}_{\hat{j}}$, and bi-fundamental fermions $( \psi_I )^{i}_{\hat{j}}$ and $( \bar{\psi}^I )^{\hat{i}}_{j}$, where $I=1,\dots,4$ is a $SU(4)$ R-symmetry index. Let us note that we are using a hat to distinguish the two sets of $U(N)$ indices. The action of the theory is
\begin{equation}
\label{ABJ(M) action}
S_{ABJ(M)}=S_{CS}+S_{Matter}+S_{int}^F+S_{int}^B \,,
\end{equation}
with
\begin{align}
S_{CS} &= -i \frac{k}{4\pi} \, \int d^3x \, \epsilon^{\mu\nu\rho} \left[ {\rm Tr} \left( A_{\mu} \partial_{\nu} A_{\rho}+ \frac{2}{3} \, A_{\mu} A_{\nu} A_{\rho} \right)- {\rm Tr} \left( \hat{A}_{\mu} \partial_{\nu} \hat{A}_{\rho}+ \frac{2}{3} \, \hat{A}_{\mu} \hat{A}_{\nu} \hat{A}_{\rho} \right) \right] \,, \nonumber \\
S_{Matter} &= \frac{k}{2\pi} \int d^3x \, \left[ {\rm Tr} \left( D_{\mu} C_I D^{\mu} \bar{C}^I \right) +i {\rm Tr} \left( \bar{\psi}^I \slashed{D} \psi_I  \right)  \right] \,, \nonumber\\
S_{int}^F &= - \frac{i k}{2\pi} \,  \int d^3x \, \left[ {\rm Tr} \left( \bar{C}^I C_I \psi_J \bar{\psi}^J \right)- {\rm Tr}\left( C_I \bar{C}^I  \bar{\psi}^J \psi_J \right) + 2 {\rm Tr} \left(C_I \bar{C}^J \bar{\psi}^I \psi_J \right) - \right. \nonumber \\
& \left. \qquad \qquad \qquad \quad  -  2 {\rm Tr} \left(\bar{C}^J C_I \psi_J \bar{\psi}^I \right) - \epsilon_{IJKL} {\rm Tr} \left( \bar{C}^I \bar{\psi}^J \bar{C}^K \bar{\psi}^L \right)  + \epsilon^{IJKL} {\rm Tr} \left( C_I \psi_J C_K \psi_L \right)  \right] \,, \nonumber\\
S_{int}^B &= - \frac{k}{6\pi} \, \int d^3x \, \left[ {\rm Tr} \left( C_I \bar{C}^I C_J \bar{C}^J C_K \bar{C}^K  \right) + {\rm Tr} \left( \bar{C}^I C_I \bar{C}^J C_J \bar{C}^K C_K  \right) + \right. \nonumber \\
& \left. \qquad \qquad \qquad \quad + 4 {\rm Tr} \left( C_I \bar{C}^J C_K \bar{C}^I C_J \bar{C}^K  \right)-6 {\rm Tr} \left( C_I \bar{C}^J C_J \bar{C}^I C_K \bar{C}^K  \right) \right] \,, \nonumber
\end{align}
where $\epsilon^{1234}=\epsilon_{1234}=1$, $k$ is the Chern-Simons level, 
and the covariant derivatives are
\begin{align}
\label{cov der-1} 
D_{\mu} C_I &= \partial_{\mu} C_I +i \left( A_{\mu} C_I -C_I \hat{A}_{\mu} \right) \,, \qquad 
D_{\mu} \bar{C}^I = \partial_{\mu} \bar{C}^I -i \left( \bar{C}^I A_{\mu} -\hat{A}_{\mu} \bar{C}^I \right) \,, \\
D_{\mu} \psi_I &= \partial_{\mu} \psi_I +i \left( \hat{A}_{\mu} \psi_I -\psi_I A_{\mu} \right) \,, \qquad \;
D_{\mu} \bar{\psi}^I = \partial_{\mu} \bar{\psi}^I 
-i \left( \bar{\psi}^I \hat{A}_{\mu} -A_{\mu} \bar{\psi}^I \right) \,.
\end{align}
In the ABJM theory the Chern-Simons level $k$ plays a the role of the inverse of a coupling constant. The 't Hooft coupling constant is defined as
\begin{equation}
\label{t hooft coupling-1}
\lambda:=\frac{N}{k} \,.
\end{equation}
In $d=3-2\epsilon$ dimensions we get that the correlators are
\begin{align}
\label{correlator-gauge field 1}
\langle \left( A_{\mu} \right)^i_j (x) \left( A_{\nu} \right)^l_m (y) \rangle &= \delta^i_m \delta^l_j \, \left( \frac{2\pi i}{k} \right) \epsilon_{\mu \nu \rho} \partial^{\rho}_x \Delta(x-y) + {\cal O} \left( \frac{1}{k^2} \right) \,, \\
\langle ( \hat{A}_{\mu} )^{\hat{i}}_{\hat{j}} (x) ( \hat{A}_{\nu} )^{\hat{l}}_{\hat{m}} (y) \rangle &= - \delta^{\hat{i}}_{\hat{m}} \delta^{\hat{l}}_{\hat{j}} \, \, \left( \frac{2\pi i}{k} \right) \epsilon_{\mu \nu \rho} \partial^{\rho}_x \Delta(x-y) + {\cal O} \left( \frac{1}{k^2} \right) \,, \\
\langle \left( C_J \right)^{\hat{i}}_{j} (x) ( \bar{C}^I)^{i}_{\hat{j}} (y) \rangle &= \frac{2\pi}{k} \, \delta^I_J \delta^{i}_j \delta^{\hat{i}}_{\hat{j}} \, \Delta(x-y) + {\cal O} \left( \frac{1}{k^2} \right) \,, \\
\label{correlator-spinor}
\langle \left( \psi_J \right)^{i}_{\hat{j}} (x) ( \bar{\psi}^I)^{\hat{i}}_{j} (y) \rangle &= \frac{2\pi}{k} \, \delta^I_J \delta^{i}_j \delta^{\hat{i}}_{\hat{j}} \, i \gamma^{\mu} \frac{\partial}{\partial x^{\mu}} \Delta(x-y)  + {\cal O} \left( \frac{1}{k^2} \right) \,,
\end{align}
where
\begin{equation}
\label{D function}
\Delta(x-y):=  \frac{\Gamma(\frac{1}{2}-\epsilon)}{4\pi^{\frac{3}{2}-\epsilon} } \frac{1}{\left( \left( x-y \right)^2 \right)^{\frac{1}{2}-\epsilon}} \,.
\end{equation}
Finally, the supersymmetry transformations of ABJM are
\begin{align}
\label{susy ABJ(M) 1}
\delta A_{\mu} &= 2 i \, \bar{\Theta}^{IJ \alpha} \left( \gamma_{\mu} \right)_{\alpha}^{\beta} \left( C_I \psi_{J \beta} + \tfrac{1}{2} \epsilon_{IJKL} \bar{\psi}^K_{\beta} \bar{C}^L \right) \,, \\
\delta \hat{A}_{\mu} &= 2 i \, \bar{\Theta}^{IJ \alpha} \left( \gamma_{\mu} \right)_{\alpha}^{\beta} \left( \psi_{J \beta} C_I + \tfrac{1}{2} \epsilon_{IJKL} \bar{C}^L \bar{\psi}^K_{\beta} \right) \,, \nonumber \\
\delta C_K &= \bar{\Theta}^{IJ \alpha} \epsilon_{IJKL} \bar{\psi}^L_{\alpha} \,, \nonumber \\
\delta \bar{C}^K &= 2 \, \bar{\Theta}^{KL \alpha} \, \psi_{L \alpha} \,, \nonumber \\
\delta \psi^{\beta}_K &= -i \bar{\Theta}^{IJ \alpha} \epsilon_{IJKL} \left( \gamma_{\mu} \right)_{\alpha}^{\beta} D_{\mu} \bar{C}^L + i \,  \bar{\Theta}^{IJ\beta} \epsilon_{IJKL} \left( \bar{C}^L C_P \bar{C}^P - \bar{C}^P C_P  \bar{C}^L \right) + \nonumber \\
& \quad \, + 2i \, \bar{\Theta}^{IJ \beta} \epsilon_{IJML} \bar{C}^M C_K \bar{C}^L \,, \nonumber \\
\label{susy ABJ(M) 6}
\delta \bar{\psi}_{\beta}^K &= -2i \,\bar{\Theta}^{KL \alpha} \left( \gamma_{\mu} \right)_{\alpha \beta} D_{\mu} C_L - 2i \, \bar{\Theta}^{KL}_{\beta} \left( C_L \bar{C}^M C_M - C_M \bar{C}^M  C_L \right) - 4 i \bar{\Theta}^{IJ}_{\beta} C_I \bar{C}^K C_J \,, \nonumber
\end{align}
where the Killing spinors $\bar{\Theta}^{IJ}$ are anti-symmetric in the R-symmetry indices ($\bar{\Theta}^{IJ} =-\bar{\Theta}^{JI}$) and satisfy the reality condition
\begin{equation}
\bar{\Theta}^{IJ}=\left( \Theta_{IJ} \right)^* \,, 
\end{equation}
with
\begin{equation}
\Theta_{IJ} =\frac{1}{2} \epsilon_{IJKL} \bar{\Theta}^{KL} \,.
\end{equation}

\section{String theory description}
\label{app: String theory description}

In this section we will propose a string dual to the $1/6$ BPS vacuum state described in \eqref{susyinsertion0}. We acknowledge collaboration of J.Aguilera-Damia for obtaining these results at the early stages of the project. 

\subsection{Type IIA string theory on $AdS_4\times {\mathbb{CP}^3}$}

The ABJM theory is conjectured to be dual to type IIA string theory on $AdS_4\times {\mathbb{CP}^3}$ \cite{Aharony:2008ug}. This background is characterized by the metric
\be
ds^2 =\frac{R^3}{4k}\left(ds^2_{AdS_4} + 4ds^2_{\mathbb{CP}^3}\right)\,,
\ee
where one can take the coordinates to be such that
\begin{alignat}{2}
ds^2_{AdS_4} & = -\cosh^2\rho \, dt^2 + d\rho^2 + \sinh^2\rho \, (d\theta^2 + \sin^2\theta d\psi^2)\,,
\\
ds^2_{\mathbb{CP}^3} & = \frac{1}{4} \left[
d\alpha^2 + \cos^2\tfrac{\alpha}{2}(d\vartheta_1^2 + \sin^2\vartheta_1 \, d\varphi_1^2)
+ \sin^2\tfrac{\alpha}{2}(d\vartheta_2^2 + \sin^2\vartheta_2 \, d\varphi_2^2)
\right.\nn\\ & \left.\qquad
+ \sin^2\tfrac{\alpha}{2}\cos^2\tfrac{\alpha}{2}(d\chi + \cos\vartheta_1 \, d\varphi_1 - \cos\vartheta_2 \, d\varphi_2)^2 \right] \,.
\end{alignat}
The other non-vanishing fields that describe the  supergravity solution are
\begin{alignat}{3}
e^{2\Phi} = \frac{R^3}{k^3}\,,\quad & F_4 = \frac{3}{8} R^3 d\Omega_{AdS_4}\,,\quad
& F_2 = \frac{k}{4} dA\,,
\end{alignat}
for
\be
 A = \cos\alpha \, d\chi +2\cos^2\tfrac{\alpha}{2}\cos\vartheta_1 \,d\varphi_1
+2\sin^2\tfrac{\alpha}{2}\cos\vartheta_2 \, d\varphi_2 \,.
\label{Abundle}
\ee

It will prove convenient to write the ${\mathbb{CP}^3}$ in terms of four complex projective coordinates $z^i$, with $i=1,\dots,4$. When restricted to $\sum_i{| z^i |^2}=1$, these complex coordinates
parametrize a 7-dimensional sphere with angles\footnote{The ranges of the angular variables are the following: $0\leq\alpha,\vartheta_1,\vartheta_2\leq\pi$, $0\leq\varphi_1,\varphi_2\leq2\pi$ and $0\leq\chi\leq4\pi$.}
 $\alpha$,$\vartheta_1$,$\varphi_1$,$\vartheta_2$,$\varphi_2$,$\chi$ and $\xi$,
according to
\be
\begin{aligned}
z^1& =\cos{\frac{\alpha}{2}}\cos{\frac{\vartheta_1}{2}}e^{\frac{i}{4}\left(2\varphi_1+\chi+\xi\right)}\,, &
z^2& =\cos{\frac{\alpha}{2}}\sin{\frac{\vartheta_1}{2}}e^{\frac{i}{4}\left(-2\varphi_1+\chi+\xi\right)}\,,\\
z^3& =\sin{\frac{\alpha}{2}}\cos{\frac{\vartheta_2}{2}}e^{\frac{i}{4}\left(2\varphi_2-\chi+\xi\right)}\,,&
z^4& =\sin{\frac{\alpha}{2}}\sin{\frac{\vartheta_2}{2}}e^{\frac{i}{4}\left(-2\varphi_2-\chi+\xi\right)}\,.
\end{aligned}
\nn
\ee
Indeed, the metric of the $\mathbb{S}^7$ can be written as a $U(1)$ bundle over $\mathbb{CP}^3$,
\be
\label{sphere coordinates}
ds^2_{\mathbb{S}^7} = ds^{2}_{\mathbb{CP}^3} + \frac{1}{16} \left(d\xi+A\right)^2 \,,
\ee
for $A$ defined in (\ref{Abundle}).

We will refer generically to the coordinates of $AdS_4\times \mathbb{CP}^3$ as $X^{m}$.
We will change from spacetime indices $m,n,\dots$ to tangent space indices $a,b,\dots$
with the following vielbein components
\be
\begin{aligned}
& e^0 = \sqrt{\tfrac{R^{3}}{4k}} \cosh\rho \, dt\,,
\quad e^1 = \sqrt{\tfrac{R^{3}}{4k}} \, d\rho \,,
\quad e^2 = \sqrt{\tfrac{R^{3}}{4k}} \sinh\rho \, d\theta\,,
\quad e^3 = \sqrt{\tfrac{R^{3}}{4k}} \sinh\rho\sin\theta \, d\psi\,,
\\
& e^4 = \sqrt{\tfrac{R^{3}}{4k}} \, d\alpha\,,
\quad e^5 = \sqrt{\tfrac{R^{3}}{4k}} \cos\tfrac{\alpha}{2} \, d\vartheta_1\,,
\quad e^6 = \sqrt{\tfrac{R^{3}}{4k}} \sin\tfrac{\alpha}{2} \, d\vartheta_2\,,
\\
& e^7 = \sqrt{\tfrac{R^{3}}{4k}} \cos\tfrac{\alpha}{2}\sin\vartheta_1 \, d\varphi_1 \,,
\quad e^8 = \sqrt{\tfrac{R^{3}}{4k}} \sin\tfrac{\alpha}{2}\sin\vartheta_2 \, d\varphi_2 \,,
\\
& e^9 = \sqrt{\tfrac{R^{3}}{4k}}  \sin\tfrac{\alpha}{2}\cos\tfrac{\alpha}{2}(d\chi + \cos\vartheta_1 \, d\varphi_1 - \cos\vartheta_2 \, d\varphi_2)\,.
\end{aligned}
\ee

The transverse scalar directions $(C_1,C_2,C_3,C_4)$ should be identified with the complex coordinates $(z^1,z^2,z^3,z^4)$. For example, the ${\rm Tr}[(C_1 \bar{C}^2)^\ell]$ vacuum is the dual operator to a string moving along the null-geodesic defined by $\rho=0$, $\alpha=0$ and
$\vartheta_1= \tfrac{\pi}{2}$.

\subsection{String with large angular momentun in $AdS_4\times \mathbb{CP}^3$ }

Let us turn now to the construction of the string solution dual to the vacuum ${\cal V}_{\ell}$  presented in \eqref{susyinsertion0}. To that aim, we should recall that for the 1/2 BPS Wilson line (\ref{general WL}) the dual open string worldsheet is an $AdS_2\subset AdS_4$ located at a fixed point in the ${\mathbb{CP}^3}$ space\cite{Drukker:2008zx}. For the choice (\ref{WLchoice}), that singles out $I=1$, we should take $|z^1| = 1$, which corresponds to put the string at $\alpha =0$ and $\vartheta_1 = 0$ in the ${\mathbb{CP}^3}$. When considering the insertion of the ${\cal V}_{\ell}$ vacuum we will generalize the ideas of \cite{Drukker:2006xg}. That is, we will consider an open string carrying a large amount of angular momentum $\ell$ in the plane 12. This configuration will be a folded string, whose folding point will follow the null-geodesic defined by $\rho=0$, $\alpha=0$ and
$\vartheta_1= \tfrac{\pi}{2}$. In that regard, we will take the ansatz
\begin{alignat}{4}
t &= \tau\,, & \qquad \quad &  & \alpha & = 0 \,, \\
\rho &= \rho(\sigma)\,, & \qquad \quad & & \vartheta_1 &=  \vartheta_1(\sigma)\,,
\\
\theta & = 0 \,, & \qquad \quad &  & \varphi_1&=  \omega\tau\,.
\end{alignat}
with all the other coordinates being zero. Our ansatz fits within a $AdS_2 \times S^2 \subset AdS_4\times {\mathbb{CP}^3}$ geometry, and therefore the classical motion
is the same as the one described in \cite{Drukker:2006xg}. More precisely, in terms of a semi-infinite worldsheet spatial parameter  $\sigma \in [0,\infty)$
 we have
\begin{align}
 \rho &= {\rm arccosh}\left(\tfrac{1}{\tanh\sigma}\right)\,,
\label{xin1}
\\
 \vartheta_1 &= \arccos\left(\tfrac{1}{\cosh \sigma}\right)\,,
\label{xin2}
\\
 \varphi_1 &= \tau\,.
\label{xin3}
\end{align}

\subsection{Supersymmetry of the folded string}

We will now discuss the supersymmetry invariance of the dual string solution proposed in the last section. In order to do so we should
first identify the Killing spinors of $AdS_4\times {\mathbb{CP}^3}$, which can be given in terms of the Killing spinors of $AdS_4\times S^7$. Using the coordinates given in \eqref{sphere coordinates} the latter 
can be written as
\be
\epsilon=\mathcal{M}\epsilon_0 \,,
\label{killesp}
\ee
with
\begin{align}
\mathcal{M} =
&
e^{\frac{\alpha}{4}(\hat{\gamma}\gamma_4-\gamma_9\gamma_{*})}
e^{\frac{\vartheta_1}{4}(\hat{\gamma}\gamma_5-\gamma_7\gamma_{*})}
e^{\frac{\vartheta_2}{4}(\gamma_{98}+\gamma_{46})}
e^{-\frac{\xi_1}{2}\hat{\gamma}\gamma_{*}}e^{-\frac{\xi_2}{2}\gamma_{57}}
e^{-\frac{\xi_3}{2}\gamma_{49}}
e^{-\frac{\xi_4}{2}\gamma_{68}}
\nn\\
&
\times
e^{\frac{\rho}{2}\hat{\gamma}\gamma_1}
e^{\frac{t}{2}\hat{\gamma}\gamma_0}
e^{\frac{\theta}{2}\gamma_{12}}e^{\frac{\psi}{2}\gamma_{23}}\,.
\end{align}
Above we have used $\xi_i$ for the phases appearing in the complex coordinates,
\be
\xi_1=\frac{2\varphi_1+\chi+\xi}{4}, \quad
\xi_2=\frac{-2\varphi_1+\chi+\xi}{4}, \quad
\xi_3=\frac{2\varphi_2-\chi+\xi}{4}, \quad
\xi_4=\frac{-2\varphi_2-\chi+\xi}{4},
\ee
and we are using the notation $\gamma_i$, $i=0, \dots ,9$, for the ten-dimensional Dirac matrices, with $\hat\gamma=\gamma_0\gamma_1\gamma_2\gamma_3$ and $\gamma_*=\prod_{i=0}^9 \gamma_i$.

When restricting to the Killing spinors of $AdS_4\times{\mathbb{CP}^3}$, we should consider only those spinors given in \eqref{killesp} that are invariant
under translations of the variable $\xi$. Under the translation $\xi\rightarrow\xi+\frac{\delta\xi}{4}$ we get
\be
\epsilon^{\prime}=\mathcal{M}e^{i\frac{\delta\xi}{8}(i\hat{\gamma}\gamma_{*}+i\gamma_{57}+i\gamma_{49}+i\gamma_{68})}\epsilon_0\,.
\label{xitrasl}
\ee
For $\epsilon'$ to be equal to $\epsilon$, we should take $\epsilon_0$ to be eigenstate of the matrices $\{ i\hat{\gamma}\gamma_{*}$, $i\gamma_{57}$, $i\gamma_{49}$, $i\gamma_{68}\}$ with eigenvalues $\{s_1,s_2,s_3,s_4\}$. Furthermore, these eigenvalues can only be $+1$ or $-1$, and they must satisfy the constraint
\be
s_1+s_2+s_3+s_4=0\,.
\label{cond}
\ee
Since in $\{s_1,s_2,s_3,s_4\}$ one can only have even numbers of $+1$ and $-1$, there are 8 combinations of eigenvalues in total.
However, the condition (\ref{cond}) rules out $\{+1,+1,+1,+1\}$ and $\{-1,-1,-1,-1\}$, and one is therefore left with
3/4 of the 32 supersymmetries,  i.e. there are 24 supersymmetries in $AdS_4\times{\mathbb{CP}^3}$.
~

As is well known, in type IIA string theory a given string configuration is supersymmetric if
\be
(1-\Gamma){\cal M}\epsilon_0 = 0\,,
\label{kappacon}
\ee
where the projector $\Gamma$ is defined as
\be
\Gamma=i\frac{\partial_\tau X^m\partial_\sigma X^n}{\sqrt{-g}}\Gamma_{mn}\gamma_{*} \,,
\label{proj}
\ee
where $g$ is the determinant of the induced metric on the worldsheet. For the family of solutions presented in (\ref{xin1})-(\ref{xin3}) we have
\be
\epsilon=e^{\frac{\vartheta_1}{4}(\hat{\gamma}\gamma_5-\gamma_7\gamma_{*})}
e^{-\frac{\tau}{4}(\hat{\gamma}\gamma_{*}-\gamma_{57})}
e^{\frac{\rho}{2}\hat{\gamma}\gamma_1}
e^{\frac{\tau}{2}\hat{\gamma}\gamma_0}\epsilon_0 \,.
 \label{killespsol}
 \ee
Moreover, the corresponding $\Gamma$ projector is
\be
\Gamma=\frac{i\left(\rho'\cosh{\rho} \,\gamma_{01} + \vartheta_1' \cosh{\rho} \, \gamma_{05}
-\rho'\sin{\vartheta_1} \, \gamma_{17} - \vartheta_1'\sin{\vartheta_1} \, \gamma_{57}\right) \, \gamma_{*}}{\sinh^2{\rho}+\cos^2{\vartheta_1}} \,,
\label{projsol}
\ee
where we have used that the solution \eqref{xin1}-\eqref{xin3} implies
\begin{equation}
\label{useful properties}
\sqrt{-g}=\sqrt{[(\rho')^2+(\vartheta_1')^2](\cosh^2 \rho-\sin^2 \vartheta_1)}=\sinh^2 \rho+ \cos^2 \vartheta_1 \,.
\end{equation}
The projector equation (\ref{kappacon}) must hold for any value of $\sigma$ and $\tau$. In particular, the only dependence on $\tau$ comes from the killing spinor (\ref{killespsol}). Since $\hat{\gamma}\gamma_{11}$ and $\gamma_{57}$ commute with
$\hat{\gamma}\gamma_{1}$, we can reshuffle factors in (\ref{killespsol}) to have
\be
\epsilon=e^{\frac{\vartheta_1}{4}(\hat{\gamma}\gamma_5-\gamma_7\gamma_{*})}
e^{\frac{\rho}{2}\hat{\gamma}\gamma_1}
e^{-\frac{\tau}{4}(\hat{\gamma}\gamma_{*}-\gamma_{57}-2\hat{\gamma}\gamma_0)}\epsilon_0\,.
\label{killespsol2}
\ee
To eliminate the $\tau$-dependence we impose the following projection condition over the constant spinor,
\be
(-i s_1 + i s_2-2\hat{\gamma}\gamma_{0})\epsilon_0 = 0\,.
\label{proj2}
\ee
Since $\hat{\gamma}\gamma_0$ does not have any zero eigenvalue, the condition (\ref{proj2}) is only satisfied if $s_1 = -s_2$. Furthermore, the constraint (\ref{proj2}) is equivalent to
\be
(1+\gamma_0\gamma_{*})\epsilon_0 =0\,, \qquad
(1-\hat{\gamma}\gamma_0\gamma_{57})\epsilon_0 =0\,.
\label{cond2}
\ee
Also, using \eqref{proj2} the equation (\ref{killespsol2}) can be rewritten as
\begin{align}
\epsilon &=
 e^{\frac{\rho}{2}\hat{\gamma}\gamma_1 +\frac{\vartheta_1}{2}\hat{\gamma}\gamma_5} \epsilon_0\,.
\end{align}
We still need to impose the kappa symmetry projection (\ref{kappacon}). Moving $e^{\frac{\rho}{2}\hat{\gamma}\gamma_1+\frac{\vartheta_1}{2}\hat{\gamma}\gamma_5}$ to the left and using \eqref{xin1}-\eqref{xin2} we get
\begin{align}
\Gamma{\cal M}\epsilon_0
= \frac{i e^{\frac{\rho}{2}\hat{\gamma}\gamma_1+\frac{\vartheta_1}{2}\hat{\gamma}\gamma_5}}{\sinh^2{\rho}+\cos^2{\vartheta_1}}
&\left[ e^{-\vartheta_1\hat{\gamma}\gamma_5}
\left(  -\sinh{\rho}\cosh{\rho}\, \gamma_{01}+\cos\vartheta_1\cosh{\rho} \,\gamma_{05}\right)+\right.\nn\\
&+\left.
e^{-\rho\hat{\gamma}\gamma_1}
\left(\sinh\rho\sin\vartheta_1\gamma_{17}-\cos\vartheta_1\sin\vartheta_1\gamma_{57}\right)\right]
\gamma_{*} \, \epsilon_0 \,,
\end{align}
and the projection (\ref{kappacon}) becomes
\begin{align}
(\sinh^2{\rho}+\cos^2{\vartheta_1}) \, \epsilon_0
=&\left[ e^{-\vartheta_1\hat{\gamma}\gamma_5}
\left(  -\sinh{\rho}\cosh{\rho} \, \gamma_{01}+\cos\vartheta_1\cosh{\rho} \, \gamma_{05}\right)+\right.\nn\\
&+\left.
e^{-\rho\hat{\gamma}\gamma_1}
\left(\sinh\rho\sin\vartheta_1\gamma_{17}-\cos\vartheta_1\sin\vartheta_1\gamma_{57}\right)\right]
\gamma_{*} \, \epsilon_0\,.
\end{align}
Expanding the exponentials and moving $\gamma_{*}$ to the left we arrive at
\begin{align}
(\sinh^2{\rho}+\cos^2{\vartheta_1}) \, \epsilon_0
=& \, \gamma_{*}\left[-\cosh{\rho}\sinh{\rho}\cos{\vartheta_1}\gamma_{01}
 -\cosh{\rho}\sinh{\rho}\sin{\vartheta_1}\hat{\gamma}\gamma_5\gamma_{01}+\right.\nn\\
 &+\left.\cosh{\rho}\cos^2{\vartheta_1}\gamma_{05}
 -\cosh{\rho} \cos{\vartheta_1} \sin{\vartheta_1} \hat{\gamma}\gamma_{0}+\right.\nn\\
&+\left.\cosh{\rho}\sinh\rho\sin\vartheta_1\gamma_{17}
+\sinh^2\rho\sin\vartheta_1\hat{\gamma}\gamma_{7}-\right.\nn\\
&-\left.\cosh\rho\cos\vartheta_1\sin\vartheta_1\gamma_{57}
-\sinh\rho\cos\vartheta_1\sin\vartheta_1\hat{\gamma}\gamma_1\gamma_{57}\right]
\epsilon_0\,,
\end{align}
which reduces, when replacing with (\ref{xin1})-(\ref{xin3}) and using (\ref{cond2}), to
\be
(1+\gamma_1)\epsilon_0=0 \,.
\ee
Therefore, taking into account \eqref{cond2} and noticing that $\gamma_1$ commutes with $\gamma_0 \gamma_*$ and $\hat{\gamma} \gamma_0 \gamma_{57}$ we conclude that the configuration preserves 4 supersymmetries, i.e. corresponds to a $1/6$ BPS solution.

\end{document}